\documentclass[lettersize, journal]{IEEEtran}
\usepackage{physics}
\usepackage{subfigure}

\usepackage{algorithm}
\usepackage{algpseudocode}
\usepackage{multirow}
\usepackage{xcolor}
\usepackage{hyperref}
\usepackage{enumitem}

\ifCLASSINFOpdf
\else
\fi
\usepackage{amsmath}
\usepackage{amssymb}
\usepackage{amsthm}
\usepackage{graphicx}
\usepackage{booktabs}

\usepackage{url}
\DeclareMathOperator*{\argmax}{argmax}

\begin{document}

\title{Multiscale Generalized Sparse Plane Wave Decomposition in the Spherical Harmonic Domain}

\author{Hüseyin Hacıhabiboğlu,~\IEEEmembership{Senior Member,~IEEE,}
\thanks{The author is with the Graduate School of Informatics, METU, Ankara, TR-06800, Türkiye (e-mail: hhuseyin@metu.edu.tr)}
\thanks{This work was supported by the Higher Education Council of T\"{u}rkiye (Y\"{O}K), Research Universities Support Programme Research Grant ADEP-704-2024-11499 ``Development of a deep-learning based MPAI standard for scene-based 6DoF immersive audio coding and compression'' (2024-2026).}
\thanks{Manuscript received March 30, 2026; revised August 06, 2026; revised October 11, 2026}
\thanks{The source code for the reference implementation of the proposed sound field analysis methods is available at \url{https://github.com/hacihabiboglu/gspwd}}}

\markboth{IEEE Transactions on Audio, Speech and Language Processing,~Vol.~XX, No.~Y, November~2026}%
{Hac\i{}habibo\u{g}lu: Multiscale Generalized Sparse Plane Wave Decomposition}

\IEEEpubid{0000--0000/00\$00.00~\copyright~2026 IEEE}

\maketitle

\begin{abstract}
Sparse plane wave decomposition (SPWD) is a physics-driven model-based approach for analyzing sound fields that is suitable for real-time operation. SPWD is used in different contexts such as direction of arrival (DOA) estimation and sound field separation. SPWD employs spherical Dirichlet kernels to decompose sound fields represented as steered response maps into a finite number of plane waves using sparse recovery techniques. This article proposes a multiscale generalized sparse plane wave decomposition approach that uses an optimal spherical kernel and a multiscale sparse decomposition method to improve recovery performance. The utility of the proposed approach is shown by comparisons with the original SPWD approach via extensive numerical simulations.
\end{abstract}

\begin{IEEEkeywords}
Sparse plane wave decomposition, spherical harmonics, orthogonal matching pursuit, multiscale sparse recovery, sound field analysis, spatial audio.
\end{IEEEkeywords}

\section{Introduction}
\IEEEPARstart{S}{ound} field analysis in the spherical harmonic (SH) domain has a variety of uses in downstream tasks such as sound source localization (e.g.~\cite{coteliMultipleSoundSource2018a, fahimMultiSourceDOAEstimation2020, lachlan2025}), acoustic source separation~\cite{varanasi2018, coteli2021sparse}, sound field separation~\cite{pezzoliSparsityBasedSoundField2022c}, sound field interpolation~\cite{Tylka_Choueiri_2020, McCormack_Politis_McKenzie_Hold_Pulkki_2022}, direct and diffuse field separation~\cite{gencDD2025}, dereverberation~\cite{peledMethodDereverberationNoise2010, wuDereverberationAlgorithmSpherical2012}, active noise control~\cite{maeno2020} and room diffuseness analysis~\cite{gotzMixingTimePrediction2015, epainSphericalHarmonicSignal2016, nolan2018}. The main purpose of sound field analysis is to obtain the directional distribution of the sound field to identify salient sources and their reflections.

Local sound fields in source-free bounded regions are 3D Herglotz, i.e.~they can be expressed as a continuous distribution of plane waves on the unit sphere~\cite{faziRepresentationSoundFields2012}. As an approximation that affords discrete solutions for sound field analysis, stationary sound fields can be expressed as a combination of a finite number of plane wave components~\cite{parkSoundfieldAnalysisPlanewave2005a}. Sound fields due to speech and music can be considered as short-time stationary when transformed under an appropriate time-frequency transform having a sufficiently fine time-frequency resolution. Therefore, it is possible to represent sound fields using a limited number of plane waves at each time-frequency box under the same time-frequency transform.

Sparse recovery techniques try to find the sparsest decomposition of a given observation vector over overcomplete dictionaries, i.e., a linear combination with the fewest non-zero coefficients that satisfies the observation constraints~\cite{crespomarquesReviewSparseRecovery2019}. {Such decompositions are useful not only in common downstream tasks such as direction of arrival (DOA) estimation, but even more importantly in problems such as sound field interpolation~\cite{olgun6DoF2026}, estimation/extrapolation~\cite{horiuchi2021, figueroaduran2025} and reconstruction~\cite{hahmann2022}.}

Sound fields can be assumed to be {approximately} sparse in the plane wave domain {for regimes dominated by direct-path and early-reflections in which the number of directionally distinct arrivals is small relative to the assumed model order~\cite{jinSoundFieldAnalysis2017}. This assumption gradually degrades as reflection density increases and when the field approaches diffuse conditions.} The complex-valued \textit{Herglotz density} on the unit sphere is the amplitude distribution that weights the plane wave components of the sound field. Each distinct plane wave appears as a Dirac impulse on the unit sphere with its complex amplitude. While the \textit{Herglotz density} provides the sparsest and sharpest representation of the sound field, it cannot be directly obtained in practice since this requires an infinite SH order.

Spherical harmonic decomposition (SHD) provides a representation that decouples direction from frequency. Rigid spherical microphone arrays (RSMAs) are typically used to obtain the SHD coefficients of the sound field~\cite{Meyer2004, Rafaely:2015do}. Since spherical microphone arrays can only provide pressure signals sampled at a finite number of points on a rigid sphere, yielding an order-limited SHD, approximating the sparse Herglotz density requires processing the raw microphone array recordings into a densely sampled spherical map followed by the application of sparse recovery techniques. 
\IEEEpubidadjcol

Beamforming in the SH domain can be used to obtain such densely sampled spherical maps of the sound field captured at an observation position~\cite{jarrettTheoryApplicationsSpherical2017b, grinsteinSteeredResponsePower2024a}. A complex-valued steered response function (SRF) can be obtained using a maximum directivity factor (maxDF) beam steered in a dense set of directions sampled on the unit sphere. Designating SRF as the observation vector and a dictionary comprising spherical Dirichlet kernels, the sound field can be represented as a sparse combination of plane waves using an approach called sparse plane wave decomposition (SPWD) (e.g.~see ~\cite{xenaki2016, coteli2021sparse}). However, the uniqueness of the identified support and the accuracy of extracted plane wave amplitudes are not guaranteed except when the sound field comprises a single dominant component since the order-limited SHD of practical spherical microphone arrays admits low-order kernels resulting in highly coherent dictionaries~\cite{olgun6DoF2026}.

This paper presents two improvements on SPWD: The first improvement involves the design of a spherical kernel that minimizes the sidelobe power which is subsequently used to obtain a beam pattern yielding generalized SRF (GSRF) maps. This facilitates the design of an associated optimal dictionary that minimizes sidelobe and backlobe power while retaining the unit-gain constraint, to improve the accuracy of the selected support. The second improvement involves a joint sparse recovery applied on multiple scales of the GSRF. This approach is based on the understanding that the support as well as the plane wave amplitudes are common across multiple GSRF scales. We demonstrate that the proposed approach performs statistically significantly better in support recovery than the original SPWD particularly at higher orders. 

The article is organized as follows. Sec.~\ref{sec:background} briefly discusses the background related to SPWD in the SH domain and identifies its limitations. Sec.~\ref{sec:gspwd} presents the improvements to SPWD as outlined above and the multiscale generalized SPWD algorithm based on multidimensional orthogonal matching pursuit. Sec.~\ref{sec:numerical} evaluates the proposed approach in terms of its accuracy. Sec.~\ref{sec:conc} concludes the article. 


\section{Background}\label{sec:background}

We will now present the conceptual background of SPWD using spherical Dirichlet kernels. While the presented content has been discussed elsewhere~\cite{coteli2021sparse} it is repeated here for motivating the proposed approach.

\subsection{Steered Response Function}

SHD of a sound field allows separating the direction and frequency dependence of sound field components. SHD coefficients are obtained via projecting the sound field in spherical coordinates onto the orthogonal basis of SH functions. For a snapshot of the sound field expressed in spherical coordinates as $p(\theta,\phi)$, the SHD coefficients, $\{p_{nm}\}$ for $n\in{}\mathbb{Z}_0^+$, $|m|\leq{}n$, can be obtained via:
\begin{equation}
    p_{nm} = \int_{S^2}p(\Omega)[Y_n^m(\Omega)]^*d\Omega
\end{equation}
with $\Omega=(\theta, \phi)$ where $\theta\in[0,\pi]$ is the zenith angle, $\phi\in[0, 2\pi)$ is the azimuth angle, and $Y_n^m(\Omega)=Y_n^m(\theta,\phi)=\sqrt{\frac{2n+1}{4\pi}\frac{(n-m)!}{(n+m)!}}P_n^m(\cos\theta)e^{im\phi}$ is the SH function {where $P_n^m(\cdot)$ is the associated Legendre polynomial} with order $n$ and degree $m$. The original distribution can be obtained back using the SHD coefficients as:
\begin{equation}
    p(\Omega)=\sum_{n=0}^{\infty}\sum_{m=-n}^{n}p_{nm}Y_n^m(\Omega)
\end{equation}

The SHD coefficients for a unit amplitude plane wave incident from $\Omega_p$ are:
\begin{equation}
    p_{nm}=\left[Y_n^m(\Omega_p)\right]^*\ \textrm{for}\ n\ge{}0\ \textrm{and}\ |m|\le{}n.
\end{equation}
Since SHD is linear, the SHD coefficients for a stationary sound field comprising $Q$ non-coincident plane waves incident from $\{(\theta_q, \phi_q)\}_{q=1\cdots{}Q}$ and with complex amplitudes $\{A_q\}_{q=1\cdots{}Q}$ can be expressed as:
\begin{equation}
    a_{nm}=\sum_{q=1}^{Q}A_q\left[Y_n^m(\Omega_q)\right]^*.\label{eq:anm}
\end{equation}
Notice that the amplitude term $A_q\in{}\mathbb{C}$ represents both the magnitude and the position-dependent phase of the plane wave.

If all SHD coefficients of the sound field are accurately known, the completeness property of the SH functions, i.e.:
\begin{equation}
    \sum_{n=0}^{\infty}\sum_{m=-n}^{n}\left[Y_n^m(\Omega_q)\right]^*Y_n^m(\Omega)=\delta(\Omega-\Omega_q)
\end{equation}
can be used to accurately extract both the directions and the amplitudes of the plane waves as Dirac impulses on the unit sphere, such that:
\begin{equation}
\sigma(\Omega)=\sum_{n=0}^{\infty}\sum_{m=-n}^{n}a_{nm}Y_n^m(\Omega)=\sum_{q}A_q\delta(\Omega-\Omega_q)
\end{equation}
where $\sigma(\Omega)$ is the Herglotz density, a complex-valued continuous distribution on the unit sphere. In other words, when the SHD is not order-limited, it is possible to obtain a representation with a support on the unit sphere, $\mathcal{Q}=\textrm{supp}(\sigma)=\{\Omega_q\}_{q=1\cdots{}Q}$, a set which contains all and only the directions of the plane waves that comprise the sound field.

In practice, three major limitations affect the calculation of the SRF. First, practical RSMAs can only provide an order-limited approximation of the SHD which is determined by the number, $M$, of microphones, such that the maximum order that can be attained, satisfies $M\ge{}(N+1)^2$, effectively limiting the highest SHD order that can be obtained without spatial aliasing\footnote{Notice that the maximum order is also dependent on the specific spherical sampling scheme, i.e.~$M\ge{}2(N+1)^2$ for Gaussian sampling~\cite{rafaely2007aliasing}.}. Second, the SRF cannot have arbitrarily high spatial resolution and has to be sampled at a finite number of points on the unit sphere. Third, mode equalization causes noise amplification at low frequencies for higher SHD orders. 

The finite order results in limiting the spatial bandwidth of the Dirac impulses. Truncating the SHD to order $N$ yields the axisymmetric and unimodal \textit{spherical Dirichlet kernel} centered at $\Omega_q$~\cite{AtkinsonHan2012}:
\begin{align}
    \Lambda^{(N)}(\Omega|\Omega_q)&=\sum_{n=0}^{N}\sum_{m=-n}^{n}\left[Y_n^m(\Omega_q)\right]^*Y_n^m(\Omega)\\&=\frac{N+1}{4\pi}P_N^{(1,0)}(\cos\Theta_q)\nonumber
\end{align}
where $\cos\Theta_q$ is the cosine angle between the directions $\Omega$ and $\Omega_q$ and $P_n^{(\alpha,\beta)}(\cdot)$ is the Jacobi polynomial of degree $n$ with shape parameters $\alpha=1$ and $\beta=0$. 

The spherical Dirichlet kernel is directly related to maxDF beamforming which maximizes the directivity index (DI) of the spatial filter~\cite{Elko2008}. MaxDF beamforming can be carried out in the SH domain by selecting the complex-valued beamforming coefficients as $Y_n^m(\Omega)$ such that the SRF can be expressed as:
\begin{equation}
    \sigma(\Omega)=\sigma^{(N)}(\Omega) + \epsilon = \sum_{q}A_q\Lambda^{(N)}(\Omega|\Omega_q)+\epsilon
\end{equation}
where $\epsilon$ is an error term that contains the higher-order terms. Observe that the order-limited portion, $\sigma^{(N)}(\Omega)$ is the Herglotz density truncated to order $N$, which is equivalent to filtering it via spherical convolution with the spherical Dirichlet kernel, producing a mainlobe of half-width $\sim\pi/(N+\frac{1}{2})$
around each source direction, surrounded by oscillatory sidelobes. 

Beamforming in a finite number of directions on the unit sphere is the second practical constraint which results in the steered response function (SRF) sampled on the unit sphere:
\begin{equation}
    s^{(N)}(\Omega_h)=\sum_{q}A_q{}\Lambda^{(N)}(\Omega_h|\Omega_q),\ \forall{}\Omega_h\in\boldsymbol{\Omega_h}
\end{equation}
where $\boldsymbol{\Omega_h}=\{\Omega_h\}_{h=0,\cdots{},H-1}\subset{}\boldsymbol{\Omega}$ with $\boldsymbol{\Omega}:=[0,\pi]\times[0,2\pi)$ are $H$ points that are uniformly or near-uniformly sampled on the unit sphere. While the value of $Q=|\mathcal{Q}|$ is usually unknown, the assumption $Q\ll{}H$ that implies sparsity is usually implicit in sound field analysis. Note that this is an idealized model not taking into account the effects of non-salient components (e.g.~late reflections and reverberation in rooms) and noise (i.e.~both spatially uncorrelated sensor noise and diffuse noise).

\subsection{Sparse Plane Wave Decomposition}

The SRF can be expressed in matrix form as:
\begin{equation}
    \mathbf{s}^{(N)}=\boldsymbol{\Phi}^{(N)}\mathbf{a}+\boldsymbol{\eta}
\end{equation}
where $\mathbf{s}^{(N)}\in\mathbb{C}^{H\times{}1}$ is the SRF vector, $\boldsymbol{\Phi}^{(N)}\in{}\mathbb{R}^{H\times{}L}$ is an overcomplete dictionary matrix containing spherical Dirichlet kernels, $\boldsymbol{\Lambda}^{(N)}_l=[\Lambda(\Omega_0|\Omega_l),\cdots{}, \Lambda(\Omega_{H-1}|\Omega_l)]^T\in{}\mathbb{R}^{H\times{}1}$ with a peak at $\Omega_l\in{}\{\Omega_l\}_{l=0\cdots{}L-1}$ such that $H\leq{}L$, sampled at the same positions, $\{\Omega_h\}_{h=0\cdots{}H-1}$ on the unit sphere, $\mathbf{a}\in{}\mathbb{C}^{L\times{}1}$ is the $Q$-sparse amplitude vector and $\boldsymbol{\eta}$ contains any residual terms as mentioned above. 

SPWD aims to identify the amplitude vector $\mathbf{a}$, given the SRF vector, $\mathbf{s}^{(N)}$. This includes the identification of both $\mathcal{S}=\textrm{supp}(\mathbf{a})$ and the amplitudes of the corresponding non-zero entries, i.e.~plane wave amplitudes. A variety of sparse recovery methods including greedy methods such as the orthogonal matching pursuit (OMP)~\cite{iwaenc2018}, $\ell_1$-minimization methods such as least absolute shrinkage and selection operator (LASSO)~\cite{routraySparsityBasedFramework2018, gencDD2025} and multiple measurement vector (MMV) models have also been previously used~\cite{wabnitz2011upscaling} in sound field analysis/synthesis contexts. 

The original SPWD approach is based on the application of OMP to obtain the least number of components that can recover the original SRF, such that:
\begin{equation}
\underset{\mathbf{a}}{\operatorname{argmin\,}} \|\mathbf{a}\|_0 \text{ subject to } ||\mathbf{s}^{(N)}-\boldsymbol{\Phi}^{(N)}\mathbf{a}||_2^2 \leq \tau    
\end{equation}
for a given tolerance $\tau$. Despite its objective to recover a representation with a parsimonious support, this formulation does not guarantee a unique solution~\cite{tropp2004}. The exact recovery conditions for sparse recovery methods including OMP had been a topic of interest over many years (e.g.~\cite{Huang_2011}). A conservative condition for OMP is based on the \textit{mutual coherence} of the employed dictionary defined as~\cite{giryesOMPHighlyCoherent2013}:
\begin{equation}
\mu(\boldsymbol{\Phi}) \triangleq \max_{i \neq j} \frac{|\langle \boldsymbol{\Phi}_i, \boldsymbol{\Phi}_j \rangle|}{\|\boldsymbol{\Phi}_i\|_2 \|\boldsymbol{\Phi}_j\|_2},\label{eq:mutcoh}  
\end{equation}
where $\boldsymbol{\Phi}_i$ is the $i$-th column of the dictionary matrix. OMP is guaranteed to recover the support of a $Q$-sparse signal exactly only if:
\begin{equation}
Q < \frac{1}{2} \left( 1 + \frac{1}{\mu(\boldsymbol{\Phi})} \right).\label{eq:appbq}
\end{equation}
It was shown using mutual coherence as a condition that the SPWD using spherical Dirichlet kernels can guarantee exact recovery only for $Q=1$ due to the high mutual coherence of the corresponding dictionary matrix at practically useful directional resolutions {and practically achievable SH orders (i.e.~for $N\leq{}6$), independently of temporal frequency}~\cite{olgun6DoF2026}.

\subsection{Key observations about SPWD}
The following key observations motivate the developments proposed in this paper:

\paragraph{Sidelobes of kernels} The spherical Dirichlet kernel maximizes the DI, which is not necessarily optimal in the sparse recovery context. More specifically, spherical Dirichlet kernel has high-magnitude sidelobe oscillations and an even higher magnitude backlobe. Both of these features but particularly the latter causes the support to be misidentified.

\paragraph{Multiscale analysis} Order-limitation of SRFs will reduce the resolution of the analysis, rendering SPWD useless in resolving multiple sources beyond the Rayleigh limit~\cite{simonsSpatiospectralConcentrationSphere2006}, particularly at low orders. When an SHD limited to $N^{\textrm{th}}$-order is available, SRFs for lower orders ($n<N$) can also be easily calculated.  SRFs at all orders will share the exact same support and coefficients. Such information redundancy can be useful in improving support recovery when used together with a \textit{hard consensus} approach across different SRF orders.

\paragraph{Parsimony vs. Accuracy trade-off:} Plane wave decomposition is commonly formulated as the recovery of the sparse Herglotz density where the support and the associated complex coefficients are jointly unknown. In practice, exact support recovery will often be ill-posed (e.g.~due to the finite angular resolution of the dictionary)\footnote{{While grid-free methods for DOA estimation problems exist that by construction eliminate basis-mismatch~\cite{xenaki2015}, their extension to spherical microphone arrays and the proposed multiscale approach is not trivial.}}.

Sparse recovery methods can potentially obtain a more accurate approximation by using more components (i.e.~producing a less sparse solution), but at the cost of identifying false positives, thereby inflating the support and inducing mismatch between the true sound scene and the solution. For many downstream tasks such as DOA estimation, a parsimonious solution is preferable to a more accurate reconstruction.

\section{Generalized Sparse Plane Wave Decomposition}\label{sec:gspwd}

We will now use the observations outlined above to improve the accuracy of sound field analysis and propose the multiscale generalized SPWD (GSPWD).

\subsection{Spherical Kernel Design}

GSPWD uses a dictionary of atoms comprising rotated versions of a real-valued axisymmetric spherical kernel that maximizes the power in a prescribed spherical cap, while suppressing the power in other directions (i.e.~sidelobes and the backlobe). More specifically, a real-valued axisymmetric spherical kernel can be expressed as a Legendre series:
\begin{equation}
    \gamma^{(N)}(\Omega)=\sum_{n=0}^{N}\beta_n^{(N)}P_n(\cos\Omega)
\end{equation}
where $\{\beta_n^{(N)}\}_{n=0\cdots{}N}$ are real-valued coefficients. Defining $\vartheta_{m}\approx{}\pi\big/(N+1/2)$ as the mainlobe half-width~\cite{simonsSpatiospectralConcentrationSphere2006}, the desired kernel is one for which the sidelobe power:
\begin{equation}
    E_{\textrm{sidelobe}}=\int_{\vartheta_m}^{\pi}[\gamma^{(N)}(\Omega)]^2\sin\Omega{}\,d\Omega
\end{equation}
is minimized with the following constraints:

\begin{align}
    \gamma^{(N)}(0)&=\sum_{n=0}^{N}\beta_n^{(N)}=1\label{eq:constraint_1}\\
    \gamma^{(N)}(\pi)&=\sum_{n=0}^{N}\beta_n^{(N)}(-1)^n=0\label{eq:constraint_2}
\end{align}
where (\ref{eq:constraint_1}) ensures that the kernel has a maximum at its axis of symmetry and (\ref{eq:constraint_2}) eliminates the backlobe. More specifically, the optimization problem can be expressed in matrix form as:
\begin{equation}
\begin{aligned}
\min_{\mathbf{b}} \quad & \mathbf{b}^\mathsf{T}\mathbf{W}\mathbf{b} \\
\text{s.t.} \quad & \mathbf{1}^\mathsf{T}\mathbf{b} = 1, \\
& \mathbf{e}^{\mathsf{T}}\mathbf{b} = 0.
\end{aligned}
\end{equation}
where $\mathbf{b}=[\beta^{(N)}_0,\cdots{},\beta^{(N)}_{N}]^{\mathsf{T}}\in\mathbb{R}^{(N+1)\times{}1}$ contains the kernel coefficients, $\mathbf{e}^{\mathsf{T}}=[(-1)^n]_{n=0}^N\in\mathbb{R}^{1\times{}(N+1)}$, and $\mathbf{W}=\big[W_{nm}\big]_{n,m=0}^{N}\in\mathbb{R}^{(N+1)\times{}(N+1)}$, such that
\begin{equation}
    W_{nm}=\int_{-1}^{\cos\vartheta_m}P_n(x)P_m(x)dx
\end{equation}
which can be numerically obtained for a given $\vartheta_m$ using Gauss-Legendre quadrature. This problem can be expressed in a more compact form as:
\begin{equation}
    \min_{\mathbf{b}} \quad  \mathbf{b}^\mathsf{T}\mathbf{W}\mathbf{b} 
\quad \text{s.t.} \quad  \mathbf{C}\mathbf{b} = \mathbf{d}.
\end{equation}
where $C=\big[\mathbf{1} | \mathbf{e} \big]^\mathsf{T}$ and $\mathbf{d}=\big[1 \ 0\big]^\mathsf{T}$, which can be identified as a quadratic program with a linear equality constraint, the solution to which can be obtained via:
\begin{equation}
    \mathbf{b}^{(N),*}=\overline{\mathbf{W}}^{-1}\mathbf{C}^{\mathsf{T}}(\mathbf{C}\overline{\mathbf{W}}^{-1}\mathbf{C}^{\mathsf{T}})^{-1}\mathbf{d}.
\end{equation}
where $\overline{\mathbf{W}}=\mathbf{W}+\lambda{}\mathbf{I}$ with $\lambda\ll{}1$, a regularization factor.

\begin{figure*}
    \centering
    \subfigure[]{\includegraphics[width=0.161\textwidth]{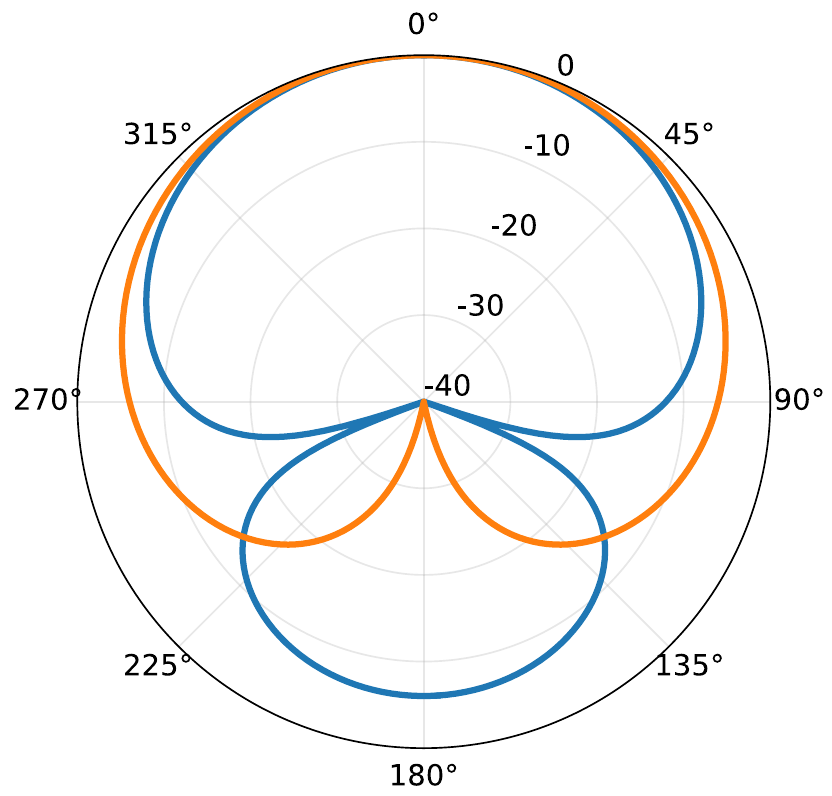}} 
    \subfigure[]{\includegraphics[width=0.161\textwidth]{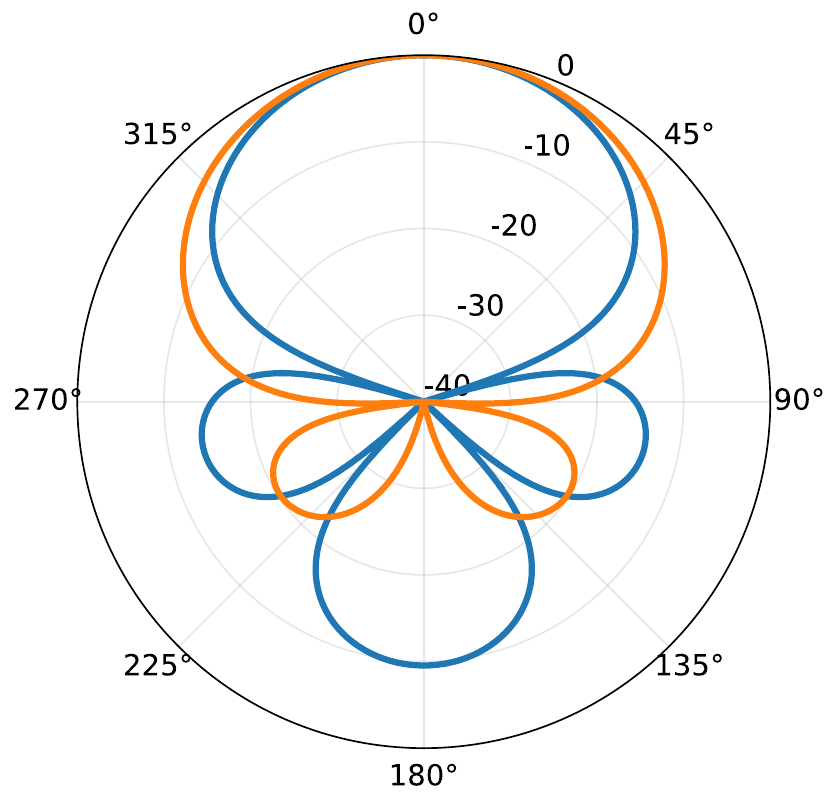}} 
    \subfigure[]{\includegraphics[width=0.161\textwidth]{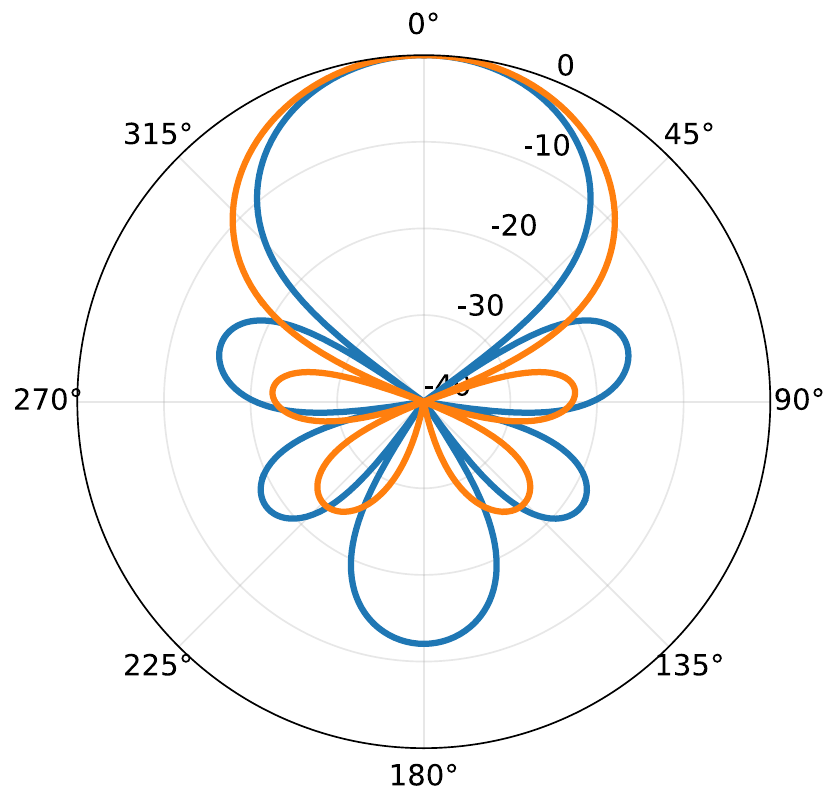}} 
    \subfigure[]{\includegraphics[width=0.161\textwidth]{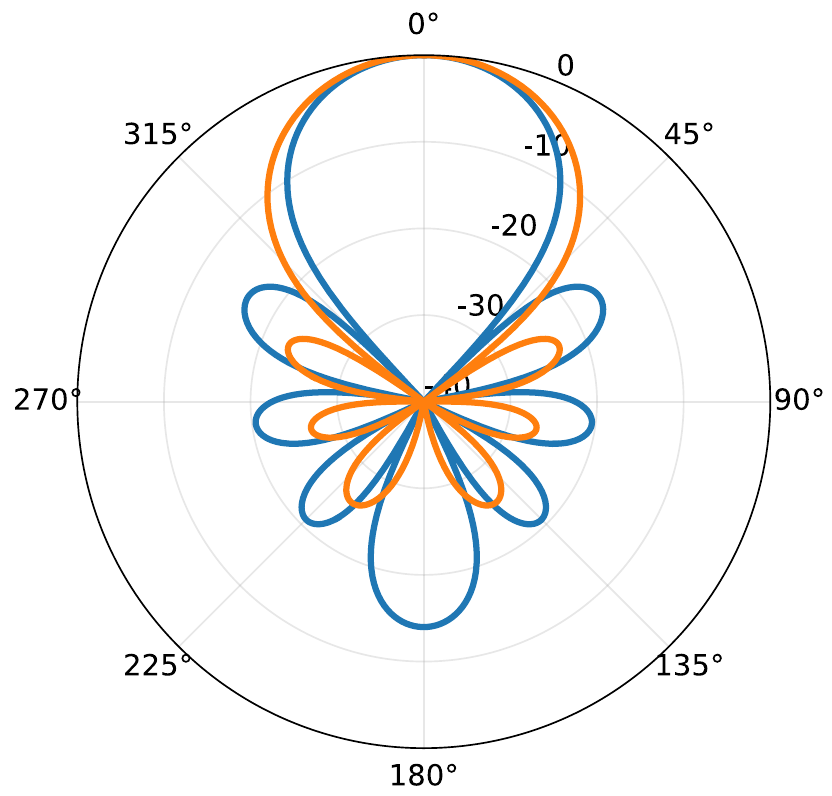}} 
    \subfigure[]{\includegraphics[width=0.161\textwidth]{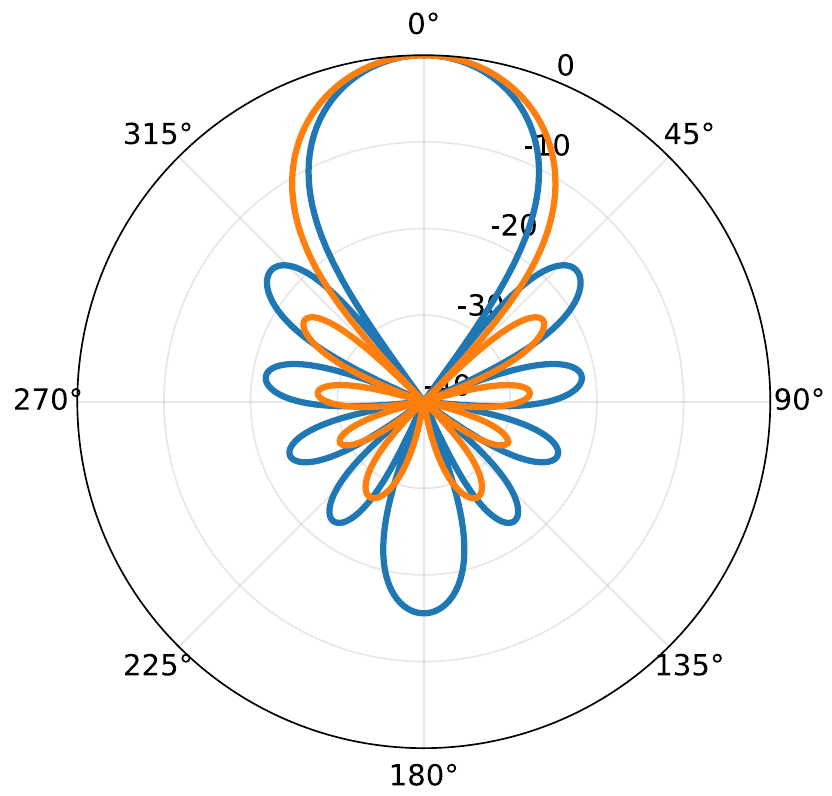}} 
    \subfigure[]{\includegraphics[width=0.161\textwidth]{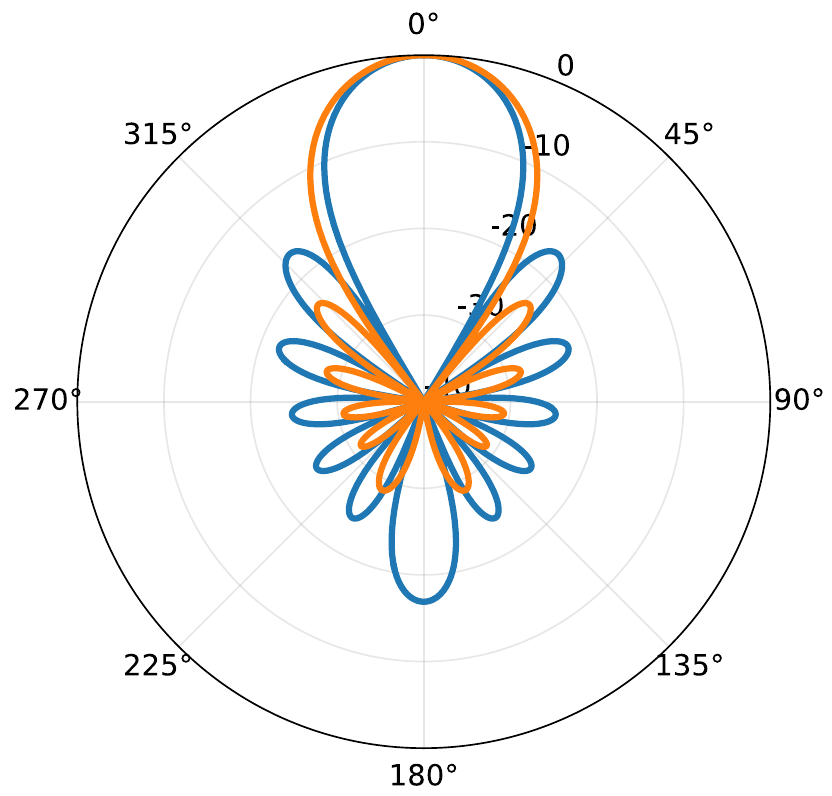}} 
    \caption{The spherical Dirichlet kernel (blue) and the proposed generalized spherical kernel (orange) for different orders, $N=1$ to $N=6$, i.e. (a)-(f), in dB. The maximum corresponds to $\Omega=0$ (i.e.~the axis of symmetry).}
    \label{fig:kernels}
\end{figure*}

Fig.~\ref{fig:kernels} shows the spherical Dirichlet kernel and the proposed kernel for $N=1\cdots{}6$. Notice that the proposed optimization approach trades mainlobe width off for sidelobe power and also explicitly sets backlobe magnitude to zero.

\subsection{Generalized SRF}
In order to obtain a spherical map that approximates Herglotz density using the spherical kernel described above, the coefficients of the latter are used as beamforming coefficients for obtaining a \textit{generalized} SRF (GSRF) such that:
\begin{equation}
    G^{(N)}(\Omega)=\sum_{n=0}^{N}\beta_n^{(N)}\sum_{m=-n}^{n}\frac{4\pi}{2n+1}a_{nm}Y_{n}^{m}(\Omega)\label{eq:gsrf}.
\end{equation}
GSRF can be sampled on the unit sphere and expressed as a linear combination of the generalized spherical kernels as:
\begin{equation}
    g^{(N)}(\Omega_h)=\sum_qA_q\gamma^{(N)}(\Omega_h|\Omega_q).
\end{equation}
\textcolor{red}{where $\gamma^{(N)}(\Omega_h|\Omega_q)$ is the value of the generalized kernel for a plane wave incident from $\Omega_q$, evaluated at a spherical grid element with a centroid direction of $\Omega_h$.}

\begin{figure}
    \centering
    \includegraphics[width=0.33\textwidth]{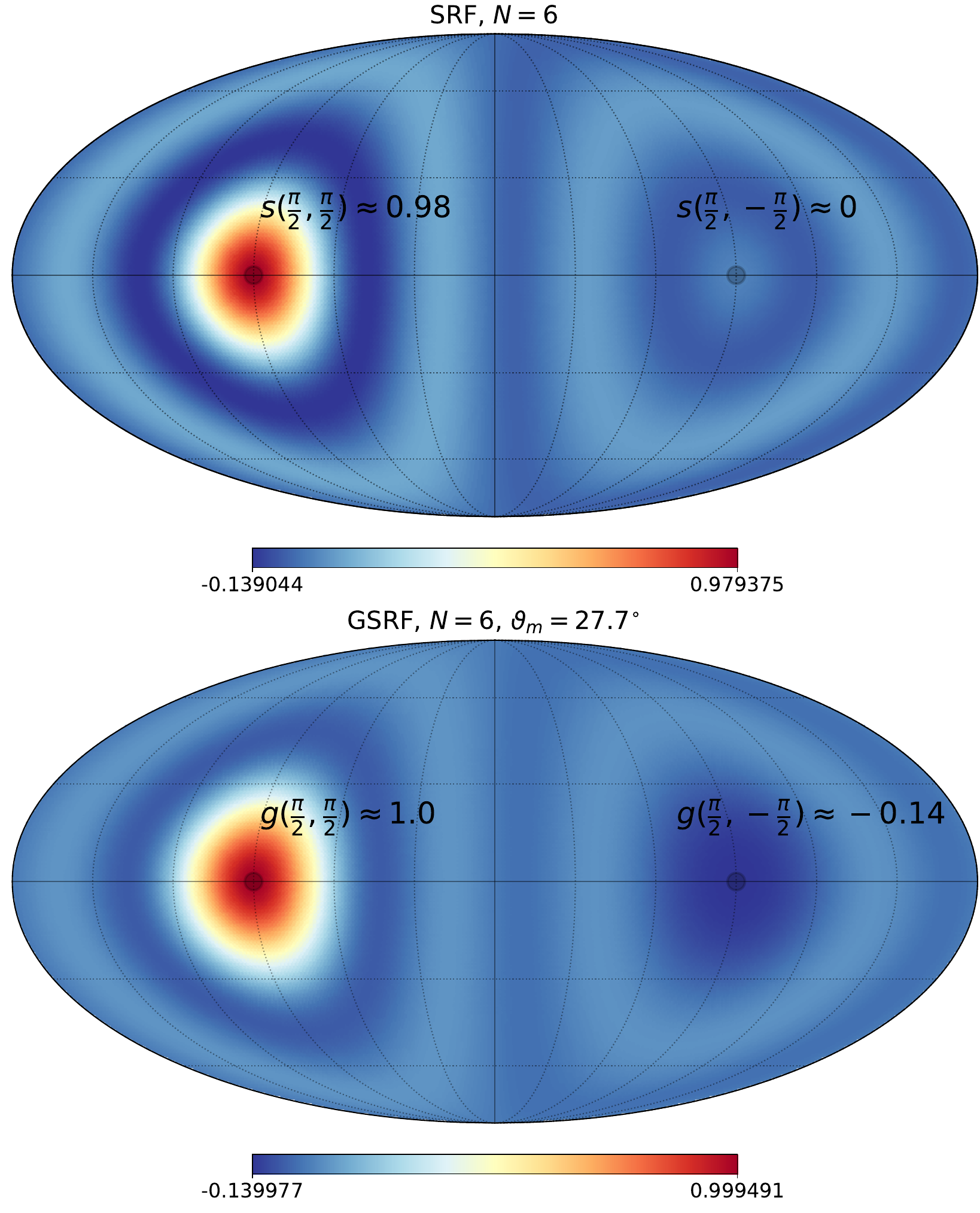}
    \caption{Comparison of SRF and GSRF of order $N=6$ for two plane waves of amplitudes $1.0$ and $-0.14$ incident from $(\frac{\pi}{2}, \frac{\pi}{2})$ and $(\frac{\pi}{2}, \frac{3\pi}{2})$, respectively. Mollweide projection is used in the plots.}
    \label{fig:srfs}
\end{figure}

Fig.~\ref{fig:srfs} shows the SRF and GSRF magnitudes of order $N=6$, for two plane waves of amplitudes $1.0$ and $-0.14$ incident from $(\frac{\pi}{2}, \frac{\pi}{2})$ and $(\frac{\pi}{2}, -\frac{\pi}{2})$, respectively, on a HEALPix grid~\cite{gorski2005healpix}. It may be observed that the peak at the direction of the second component vanishes in SRF, while being correctly represented in GSRF.

\subsection{Multiscale GSPWD}
For each order, $N$, the GSRF can be expressed in matrix form as:
\begin{equation}
\mathbf{g}^{(N)}=\boldsymbol{\Psi}^{(N)}\mathbf{a}+\boldsymbol{\eta}
\end{equation}
where $\mathbf{g}^{(N)}=[g^{(N)}(\Omega_0),\cdots{},g^{(N)}(\Omega_{H-1})]^\mathsf{T}\in{}\mathbb{C}^{H\times{}1}$ contains samples of the GSRF evaluated on a spherical grid at $\{\Omega_h\}_{h=0\cdots{}H-1}$, $\boldsymbol{\Psi}^{(N)}=\big[\boldsymbol{\Gamma}_0^{(N)}, \cdots,\ \boldsymbol{\Gamma}_{L-1}^{(N)}\big]\in{}\mathbb{R}^{H\times{}L}$ with $\boldsymbol{\Gamma}_{l}^{(N)}=\big[\gamma^{(N)}(\Omega_0|\Omega_l),\cdots{}\ \gamma^{(N)}(\Omega_{H-1}|\Omega_{l})\big]^{\mathsf{T}}\in{}\mathbb{R}^{H\times{}1}$ where $L\geq{}H$ is the number of pixels in the analysis grid and $H$ is the grid over which the $N^{\mathrm{th}}$-order GSRF is calculated. Notice that $\mathrm{rank}\big(\boldsymbol{\Psi}^{(N)}\big)=(N+1)^2$ for $H\geq(N+1)^2$ since the spherical kernels are linear combinations of SH functions up to order $N$. 

Multiscale GSPWD is based on the joint selection of atoms using each GSRF order. This is similar to the MMV model~\cite{chenSparseRepresentationsMultiple2005}, except in the latter,  the measurement vectors share the same support in the \textit{same} dictionary but with different activations, while in the present problem, the sparsity profile as well as the activation amplitudes are the same across different dictionaries as the plane wave amplitude vector, $\mathbf{a}$ is order-invariant. The same problem can also be posed as a \textit{stacked} single measurement vector (SMV) problem, but with a different selection component criterion based on hard consensus as described below.

\subsubsection{Pursuit}
The pursuit stage involves the selection of atoms that provide a good fit to data. In order to account for GSRFs at multiple scales, this stage involves hard consensus enforced via harmonic mean (i.e.~$\ell_{-1}$-norm) of inner products with atoms. This selection rule penalizes per-order mismatch, ensuring that the selected atoms are uniformly consistent across all SRF orders. Let us define the atom selection criterion at iteration, $t$:
\begin{equation}
    l_t^{*}=\argmax_l{}N{}\left[\sum_{n=1}^{N}\left(\frac{\left|\left\langle \boldsymbol{\Gamma}^{(n)}_l, \mathbf{r}_{t}^{(n)} \right\rangle\right|}{\max_{l'}\left|\left\langle \boldsymbol{\Gamma}^{(n)}_{l'},\, \mathbf{r}_{t}^{(n)} \right\rangle\right|}+\textrm{eps}\right)^{-1}\right]^{-1}
\end{equation}
where $\mathbf{r}_{0}^{(n)}=\mathbf{g}^{(n)}$, {$\textrm{eps}$ is the machine epsilon} and the residual at iteration $t+1$ is calculated as:
\begin{equation}
    \mathbf{r}_{t+1}^{(n)}=\left[\mathbf{I}-\boldsymbol{\Psi}_{t}^{(n)}(\boldsymbol{\Psi}_{t}^{(n),\mathsf{T}}\boldsymbol{\Psi}_{t}^{(n)})^{-1}\boldsymbol{\Psi}_{t}^{(n),\mathsf{T}}\right]\mathbf{g}^{(n)}
\end{equation}
where $\boldsymbol{\Psi}_{t}^{(n)}=\boldsymbol{\Psi}^{(n)}\mathbf{E
}_t\in{}\mathbb{R}^{H\times{}(t+1)}$ and the atom selection matrix defined as:
\begin{equation}
\mathbf{E}_t=[\mathbf{e}_{0},\ \mathbf{e}_{1},\ \cdots,\ \mathbf{e}_{t}]\in{}\mathbb{R}^{L\times{}(t+1)}    
\end{equation}
with $\mathbf{e
}_t=[\mathbf{e}_t]_k=\delta_{kl_t^{*}}\in{}\mathbb{R}^{L\times{}1}$ and the support $\mathcal{S}$ are updated at each iteration.

Notice that if an atom has a very low fit at a given order the objective will be very small. This ensures that only the atom indices that simultaneously provide a good fit at all GSRF orders are selected.

\subsubsection{Orthogonal Projection}
Orthogonal projection involves obtaining a residual vector that is orthogonal to the span of the already selected atoms. As in single measurement vector OMP, the iterations start with:
\begin{equation}
    \mathbf{R}_0=\mathbf{G}
\end{equation}
where $\mathbf{G}=\left[\mathbf{g}^{(1),\mathsf{T}}, \mathbf{g}^{(2),\mathsf{T}}\cdots\mathbf{g}^{(N),\mathsf{T}}\right]^{\mathsf{T}}\in{}\mathbb{C}^{NH\times{}1}$. The dictionary matrices can be stacked to obtain the multiscale dictionary matrix, $\mathbf{D}=[\boldsymbol{\Psi}^{(1), \mathsf{T}}, \boldsymbol{\Psi}^{(2), \mathsf{T}},\cdots{},\boldsymbol{\Psi}^{(N), \mathsf{T}}]^\mathsf{T}\in{}\mathbb{R}^{NH\times{}L}$ and the matrix of selected atoms at iteration $t$ is given as $\mathbf{D}_t=\mathbf{D}\mathbf{E}_t\in{}\mathbb{R}^{NH\times{}(t+1)}$. The \textit{global} residual is then calculated at each iteration as:
\begin{equation}
    \mathbf{R}_{t+1}=\left[\mathbf{I}-\boldsymbol{D}_{t}\left(\boldsymbol{D}_{t}^{\mathsf{T}}\boldsymbol{D}_{t}\right)^{-1}\boldsymbol{D}_{t}^{\mathsf{T}}\right]\mathbf{G}.
\end{equation}

The stopping condition is either after a prescribed number of iterations for a known sparsity level such that $|\mathcal{S}|=Q$, or when the relative residual energy is below a specified threshold, e.g.
\begin{equation}
    \frac{1}{N}\left({\|\mathbf{R}\|_F^2}\Big/{\|\mathbf{G}\|_F^2}\right)\leq{}\textrm{THR}
\end{equation}
where $\|\cdot\|_F$ and $\|\cdot\|_2$ represent the Frobenius and Euclidean norms, respectively. Algorithm~\ref{alg:md-ss-omp} shows the pseudocode for the proposed GSPWD algorithm using the latter stopping rule. {Note that it may in principle be possible to employ a statistically motivated model selection approach such as the deviance information criterion (DIC) as in~\cite{nolan2023} to explicitly select the model sparsity but this is not the approach taken in here.}

\begin{algorithm}[t]
\caption{Multiscale Generalized SPWD}\label{alg:md-ss-omp}
\begin{algorithmic}[1]
\setlength{\itemsep}{2pt}
\Require GSRFs $\{\mathbf{g}^{(n)} \in \mathbb{C}^{H}\}_{n=1}^{N}$,
         dictionaries $\{\boldsymbol{\Psi}^{(n)} \in \mathbb{R}^{H \times L}\}_{n=1}^{N}$,
         threshold $0 < \mathrm{THR} \ll{} 1$
\Ensure Coefficient vector $\mathbf{a} \in \mathbb{C}^{L}$ with $\textrm{supp}(\mathbf{a})=\mathcal{S} \subseteq \{1,\ldots,L\}$, $0<|\mathcal{S}|\ll{}L$

\State $\mathbf{r}_0^{(n)} \leftarrow \mathbf{g}^{(n)}$ for all $n$

\State $\mathbf{G} \leftarrow \bigl[\mathbf{g}^{(1)\mathsf{T}}, \ldots, \mathbf{g}^{(N)\mathsf{T}}\bigr]^{\mathsf{T}}$,\quad
       $\mathbf{D} \leftarrow \bigl[\boldsymbol{\Psi}^{(1)\mathsf{T}}, \ldots, \boldsymbol{\Psi}^{(N)\mathsf{T}}\bigr]^{\mathsf{T}}$
\State $\mathcal{S} \leftarrow \emptyset$,\quad $t \leftarrow 0$,\quad 
\State $\mathbf{E}_{-1}\leftarrow{}\left[\ \right]$

\Repeat
    \For{$n = 1, \ldots, N$}
        \State $c_{n,l} \leftarrow
               \left|\left\langle \boldsymbol{\Gamma}^{(n)}_l,\, \mathbf{r}_{t}^{(n)} \right\rangle\right|
               \Big/
               \max_{l'}\left|\left\langle \boldsymbol{\Gamma}^{(n)}_{l'},\, \mathbf{r}_{t}^{(n)} \right\rangle\right|$
    \EndFor

        \State $\mathrm{score}_l \leftarrow
               N\!\left[\displaystyle\sum_{n=1}^{N}
               (c_{n,l}+\textrm{eps})^{-1}\right]^{-1}$
               \quad for $l = 1,\ldots,L$
        
    \State $l_t^{*} \leftarrow \displaystyle\argmax_{l}\;\mathrm{score}_l$
    \State $\mathbf{E}_t \leftarrow [\mathbf{E}_{t-1}\big|\mathbf{e}_{l_t^*}]$
       \Comment{Atom selection matrix}
    \State $\mathcal{S} \leftarrow \mathcal{S} \cup \{l_t^{*}\}$,\quad $\mathbf{D}_t \leftarrow \mathbf{D}\mathbf{E}_t$
    \For{$n = 1, \ldots, N$}
        \Comment{Per resolution orth. proj.}
        \State $\boldsymbol{\Psi}_t^{(n)} \leftarrow \boldsymbol{\Psi}^{(n)}\mathbf{E}_t$
        \State $\mathbf{r}_{t+1}^{(n)} \leftarrow
               \left[\mathbf{I} -
               \boldsymbol{\Psi}_t^{(n)}
               \!\left(\boldsymbol{\Psi}_t^{(n)\mathsf{T}}\boldsymbol{\Psi}_t^{(n)}\right)^{-1}\!
               \boldsymbol{\Psi}_t^{(n)\mathsf{T}}
               \right]\mathbf{g}^{(n)}$
    \EndFor

    \State $\mathbf{a} \leftarrow
           \arg\min_{\mathbf{a}}\,\left\|\mathbf{G} - \mathbf{D}_t\,\mathbf{a}\right\|_2$
           \Comment{Global least-squares}
    \State $\mathbf{R}_{t+1} \leftarrow
           \left[\mathbf{I} -\mathbf{D}_t\!\left(\mathbf{D}_t^{\mathsf{T}}\mathbf{D}_t\right)^{-1}\!\mathbf{D}_t^{\mathsf{T}}
           \right]\mathbf{G}$

    \State $t \leftarrow t + 1$

\Until{$\dfrac{1}{N}\!\left({\|\mathbf{R}_t\|_F^2}\big/{\|\mathbf{G}\|_F^2}\right) \leq \mathrm{THR}$}

\State \Return $\mathbf{a},\;\mathcal{S}$
\end{algorithmic}
\end{algorithm}

\section{Numerical Evaluation}\label{sec:numerical}

Numerical simulations were run in order to assess the performance of the proposed method. We compare four methods: 1) single SPWD, 2) single GSPWD both of which use only the highest-order SRF and GSRF, respectively, 3) multiscale SPWD and 4) multiscale GSPWD that use all SRFs and GSRFs, respectively, with all orders from $N=1$ to $N=6$.

\subsection{Evaluation Measures}
We use {three} direction-thresholded measures and a support recovery measure to assess the performances of the methods. Before the calculation of these measures, the support obtained from the compared methods are mapped to the ground truth via linear sum assignment~\cite{hungarian1955}. More specifically, the angular distance between unit vectors, $\mathbf{n}_p$ and $\mathbf{n}_q$ representing two directions is given by:
\begin{equation}
d_{(p,q)}=\cos^{-1}\left\langle\mathbf{n}_p,\mathbf{n}_q\right\rangle\ \forall{}(p,q)\in\mathcal{L}
\end{equation}
Linear sum assignment produces the set of tuples $\mathcal{L}=\{(p,q)\ \big|\ p\in\mathcal{S}, q\in\mathcal{Q}\}$ that minimizes the total assignment cost, i.e.~the total angular distance. Notice that the atom direction $\mathbf{n}_p$ is sampled on a discrete grid defined on the unit sphere, while the ground truth direction $\mathbf{n}_q$ comes from a continuous distribution. The employed performance measures are:

\subsubsection{F1-score (F1)}

The direction-thresholded F1-score~\cite{politisOverviewEvaluationSound2021a} is used to assess the predictive performance of the compared methods. This involves calculating the direction-thresholded errors such that:
\begin{align}
     &TP_{\leq{}\Theta} =\sum_{p\in\mathcal{S}}1_{d\leq{}\Theta}  \\
     &FP =\max(0, |\mathcal{S}|-|\mathcal{Q}|)+\min(|\mathcal{S}|, |\mathcal{Q}|) - TP_{\leq{}\Theta}\\
     &FN =|\mathcal{Q}|-TP_{\leq{}\Theta}
\end{align}
where $1_{d\leq{}\Theta}=1$ when $d\leq{}\Theta$ and zero otherwise and terms defined above are the \textit{true positive}, \textit{false positive} and \textit{false negative}, respectively. The direction-thresholded F1-score is calculated from these error terms as per its common definition\footnote{The direction-thresholded F1-score employed here is different from the one given in~\cite{politisOverviewEvaluationSound2021a} in which a component matched to a ground truth source but with an angular distance greater than the threshold are ignored in the calculation of FN.}.

\subsubsection{Event Count Mismatch (ECM)} The number of plane waves in the ground truth, $|\mathcal{Q}|$, and the number of identified plane waves, $|\mathcal{S}|$, are used for calculating the event count mismatch, defined as: 
{
\begin{equation}
    \textrm{ECM}={\big||\mathcal{S}|-|\mathcal{Q}|\big|}\big/{|\mathcal{Q}|},
\end{equation}
}
where a value close to zero indicates a better match. Notice that while ECM  resembles the event count recall (ECR) measure defined in~\cite{politisOverviewEvaluationSound2021a} it is different in the sense that any deviation from the ground truth in terms of the number of terms in the solution is equally penalized unlike ECR which is a binary measure of mismatch. 

We also report two other derived measures:
\begin{enumerate}[label=\alph*)]
    \item \textit{ECM Rate} comprises three terms indicating the ratio of over-selection, under-selection and the exact selection to the total number of simulations in a specific scenario. 
    \item \textit{ECM Magnitude} comprises the number of over- and under-selected components and indicates the severity of mismatch in terms of the number of atoms used. The signed ECM Magnitude indicates the direction of mismatch (i.e.~negative values indicate under-selection and vice versa). While ECM and ECM Rate are ratios and are therefore unitless, ECM Magnitude is an absolute measure in atoms.
\end{enumerate}

\subsubsection{Support-aware Amplitude Error (SAAE)} The error between the matched components as well as the spurious and missed components are used to calculate the support-aware amplitude error, which is defined as:
{\begin{equation}
SAAE_{\leq{}\Theta}=10\log_{10}\left(\frac{E_{\leq{}\Theta}+E_{\textrm{missed}}+E_{\textrm{spurious}}}{E_{\textrm{true}}}\right)
\end{equation}}
{where the four terms that comprise SAAE are defined as:}
{\begin{align}
    E_{\leq{}\Theta} &= \sum_{(p,q)\in{}\mathcal{L}}1_{d_{(p,q)}\leq{}\Theta}\ |A_p-A_q|^2,\\
    E_{\textrm{missed}} &= \sum_{q\in\mathcal{Q}}|A_q|^2\left(1-\sum_{p|(p,q)\in{}\mathcal{L}}1_{d_{(p,q)}\leq{}\Theta}\right),\\
    E_{\textrm{spurious}} &= \sum_{p\in\mathcal{S}}|A_p|^2\left(1-\sum_{q|(p,q)\in{}\mathcal{L}}1_{d_{(p,q)}\leq{}\Theta}\right),\\
    E_{\textrm{true}} &= \sum_{q\in\mathcal{Q}}|A_q|^2.
\end{align}}
Here, $E_{\leq{}\Theta}$ is the total direction-thresholded squared error over matched components, $E_{\textrm{missed}}$ is the total squared error due to the components that were not identified by the corresponding method, $E_{\textrm{spurious}}$ is the total squared error due to the spuriously identified components, and $E_{\textrm{true}}$ is the total energy of the individual components.

\subsubsection{Localization Error (LE)} 
Localization error (LE) is the angular distance between the identified directions and the ground truth, averaged over the total number of identified directions. Since the proposed methods have finite spatial precision due to the resolution of the spherical maps, a direction threshold, $\Theta$, is also used in the calculation of localization error. The direction-thresholded LE is defined as~\cite{politisOverviewEvaluationSound2021a}:
\begin{equation}
    LE_{\leq{}\Theta} = \frac{\sum_{(p,q)\in{}\mathcal{L}}1_{d_{(p,q)}\leq{}\Theta}\ d_{(p,q)}}{\sum_{(p,q)\in{}\mathcal{L}}1_{d_{(p,q)}\leq{}\Theta}+\textrm{eps}}.
\end{equation}

Notice, however, that LE by itself can be misleading in different ways. \textcolor{red}{For example, a method can identify a very small number of sources (or none), but with good accuracy leading to a low (or vanishing) LE.} Similarly, a method that identifies a high number of components can have a lower LE than one that provides a sparser solution since each additional component increases the directional \textit{coverage} on the unit sphere.

\subsection{Simulation Setup}

We simulated random sound scene scenarios by synthetically generating spherical maps with varying number of components, noise and SH order. In order to achieve this, we use synthetically generated SH decomposition vectors and carry out modal beamforming to obtain the steered response maps using the spherical Dirichlet and generalized kernel coefficients as described above. 

Simulated SH coefficients were calculated for different numbers of plane waves, i.e.~up to and including $10$, and different SNR levels ($10$, $20$, $30$, $40$, $\infty$ dB) up to $N_{\max}\leq{}6$, an SH order that is practically attainable by state-of-the-art spherical microphone arrays such as the Eigenmike em64~\cite{mhacoustics2023em64}. 

{The signal model for the simulated scenes is given as:
\begin{equation}
    \tilde{a}_{nm}=\sum_{q}\tilde{A}_q{}\left[Y_{n}^{m}(\tilde{\Omega}_q)\right]^*+\tilde{\epsilon}_{nm}
\end{equation}
for $n\leq{}N_{\max}$ and $|m|\leq{}n$ where $\tilde{A}_q{}$, $\tilde{\Omega}_q$, and $\tilde{\epsilon}_{nm}$ represent the amplitude and direction of a single component and the noise term, respectively.}

{Rejection sampling was used to obtain the source directions for each trial. More specifically, SH coefficients of the simulated scene are calculated as a linear sum as in (\ref{eq:anm}) with the incidence angles $\{\tilde{\Omega}_q=(\tilde{\theta}_q, \tilde{\phi}_q)\;|\;\tilde{\theta}_q\thicksim\arccos{(1-2u_{\theta}}), \tilde{\phi}_q\thicksim 2\pi{}u_{\phi}\}$ such that $u_{\theta}, u_{\phi}\thicksim{}U(0,1)$ where $U(0,1)$ is the uniform distribution. Rejection sampling involves sequential selection of directions as follows. Given an already accepted set $\mathcal{Q}_{q-1}=\{\Omega_1,\cdots,{}\Omega_{q-1}\}$ the next randomly drawn candidate $\Omega^{*}$ is accepted based on:}
\begin{equation}
    \textrm{accept}\ \Omega^{*} \iff \min_{\Omega_j\in{\mathcal{Q}_{q-1}}}\Theta_{j,*}\geq{}\delta
\end{equation}
where $\Theta_{j,*}$ is the angle between the two unit vectors coincident with the directions $\Omega_j$ and $\Omega_*$ and $\delta=\pi/\max(3N_{\max}, 4)$ is the rejection threshold with $N_{\max}$ being the maximum SH order employed in that specific scenario. This selection results in a directional resolution less than half the Rayleigh limit~\cite{simonsSpatiospectralConcentrationSphere2006} in all tested orders.\footnote{Notice that calculation of the SH coefficients of the scene directly in the SHD domain represents an ideal case with no spatial aliasing due to sampling.}

{The amplitudes of the plane wave components whose directions are selected as above were sampled from the complex normal distribution, i.e.~ $\tilde{A}_q\thicksim{}\mathcal{CN}(0,1)$. A randomly selected amplitude was clamped via scaling such that $|A_q|\xleftarrow{}\max(0.2, \min(|A_q|, 1))$ corresponding to approximately $14$ dB difference between the most and least prominent components in the worst case. Finally, once the SH coefficients for the scene were calculated, (generalized) SRFs were obtained for a given order $N$ from (\ref{eq:gsrf})}.

{Additive noise used in the simulations was obtained directly in the SH domain. This was done by drawing the noise SH coefficients $\left\{\tilde{\epsilon}_{nm}\right\}$ from the complex normal distribution $\mathcal{CN}(0,1)$ so that the noise is isotropic (i.e.~its spherical spectrum is statistically flat). In order to ensure that the plane wave with the lowest amplitude stays above the noise floor, SNR was defined as the ratio of the power of the lowest component to the noise such that:}

\begin{equation}
    \textrm{SNR}=10\log_{10}\frac{\sum_{n,m}\left|\check{A}\left[Y_{n}^{m}(\check{\Omega})\right]^*\right|^2}{\sum_{n,m}|\epsilon_{nm}|^2}\ (dB)
\end{equation}
where $\check{A}$ and $\check{\Omega}$ are the amplitude and direction of the component with the minimum magnitude in a specific scene, respectively. Each test condition was repeated $50$ times using randomly generated directions and complex amplitudes. This way, while the noise power changes from simulation to simulation according to the minimum amplitude in the scene, the scene-dependent SNR stays the same.

\textcolor{red}{For every combination of SH order and source count in the evaluation, the random scene generator was initialized with the same seed, so that the same $50$ scenes were evaluated at different SNR levels. Since the minimum angular separation, $\delta$, depends on $N_{\max}$, the scenes for different SH orders were drawn separately. Due to the common seed, scenes for different choices of SH order and source count are not strictly independent and may share some source directions. Overall, $3000$ distinct scenes ($10$ source counts $\times$ $6$ SH orders $\times$ $50$ repetitions) were each evaluated at $5$ SNR levels, totalling $15000$ trials.} For each randomly generated scene, SRF and GSRF vectors were calculated up to order $N=6$ on a $48$ pixel HEALPix grid which has a resolution of $29.32^{\circ}$. Dictionary matrices were calculated using a $3072$-pixel HEALPix grid that provides a resolution of $3.66^{\circ}$.

The threshold used as the stopping criterion was selected as $10^{-2}$ which corresponds to a relative residual power of $-20$ dB for the single-scale methods. While calculating the evaluation measures, the angular threshold was selected as $29.32^{\circ}$ (i.e.~the resolution of the 48-pixel grid). The statistical analyses were carried out using JASP software (version 0.98.1)~\cite{JASP2025}.

\subsection{Results}

\subsubsection{F1-score (F1)}

\begin{figure}
    \centering
    \includegraphics[width=0.4\textwidth]{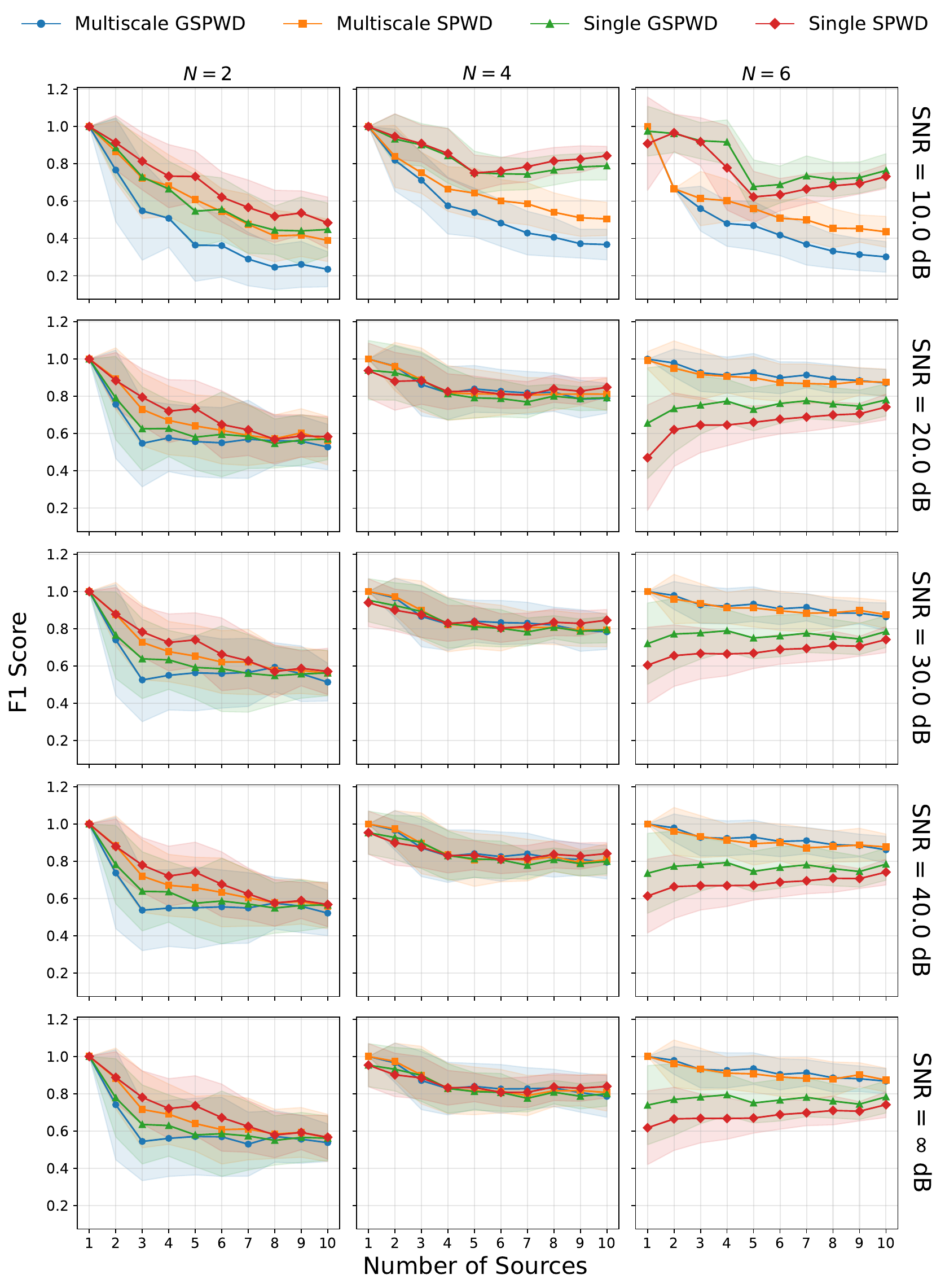}
    \caption{F1-scores for different methods and number of sources. Only the conditions for $N=2$, $N=4$ and $N=6$ are shown at $10$, $20$, $30$, $40$ and $\infty$ dB SNR. The curves represent the mean values and shaded areas represent one standard deviation above and below the mean.}
    \label{fig:f1}
\end{figure}

Part of the F1-score results for the tested methods are shown in Fig.~\ref{fig:f1}. The results indicate that while {single scale SPWD} provides a better performance at low SH orders (e.g.~for $N=2$), increasing the SH order increases the performances of multiscale versions of both SPWD and GSPWD. \textcolor{red}{Multiscale methods outperform the single-scale methods particularly at the highest SH orders in the evaluation (i.e.~$N=5$ and $6$) while achieving an F1-score performances that are close to ideal except when $\textrm{SNR}=10$ dB.}

\textcolor{red}{A repeated-measures ANOVA (rmANOVA) of F1-score with \textit{Method} and \textit{SNR} as within-subjects factors and \textit{SH Order} and \textit{Number of Sources} as between-subjects factors was carried out. Since the sphericity assumption was not satisfied for \textit{Method}, \textit{SNR} and their interaction (Mauchly's test, $p < 0.001$), Greenhouse-Geisser correction was applied to the degrees of freedom with $\varepsilon = 0.848$, $0.408$ and $0.490$, respectively.}

\textcolor{red}{The rmANOVA indicated that all of the main effects and interactions were statistically significant ($p < 0.001$). \textit{SH Order} [$F(5, 2940) = 1513.324$, $\omega^2 = 0.300$] had a large effect size, while \textit{SNR} [$F(1.632, 4799.508) = 2586.945$, $\omega^2 = 0.136$] and \textit{Number of Sources} [$F(9, 2940) = 362.839$, $\omega^2 = 0.100$] had medium effect sizes. While \textit{Method} [$F(2.543, 7477.638) = 71.076$, $\omega^2 = 0.009$] had a negligible effect size, its interaction with \textit{SNR} [$F(5.884, 17298.825) = 2096.733$, $\omega^2 = 0.137$] had a medium effect size, and its interaction with \textit{SH Order} [$F(12.717, 7477.638) = 123.918$, $\omega^2 = 0.031$] as well as the \textit{Method} $\times$ \textit{SNR} $\times$ \textit{SH Order} interaction [$F(29.420, 17298.825) = 190.873$, $\omega^2 = 0.037$] had small effect sizes which indicate that the different F1-score performances between the methods depend on SNR and the SH order. \textit{SNR} $\times$ \textit{Number of Sources} [$F(14.692, 4799.508) = 151.200$, $\omega^2 = 0.033$], \textit{SNR} $\times$ \textit{SH Order} [$F(8.162, 4799.508) = 71.208$, $\omega^2 = 0.012$] and \textit{SH Order} $\times$ \textit{Number of Sources} [$F(45, 2940) = 31.184$, $\omega^2 = 0.010$] also had small effect sizes.}


\begin{figure}[!t]
    \centering
    \includegraphics[width=0.4\textwidth]{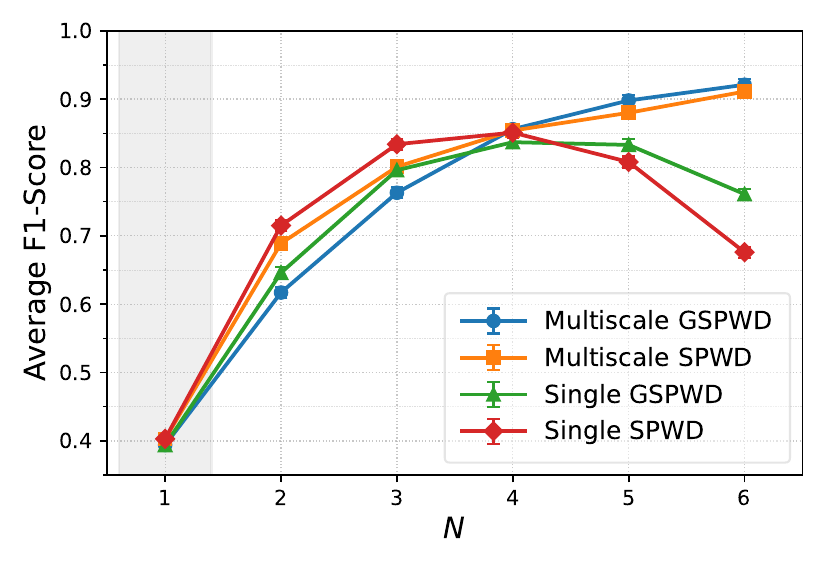}
    \caption{Average F1-scores for different methods at different SH orders showing single SPWD (red diamonds), single GSPWD (green triangles), multiscale SPWD (orange squares), and multiscale GSPWD (blue circles). $N=1$ for which multiscale versions collapse into their single scale versions is indicated by gray shading.}
    \label{fig:f1marginal}
\end{figure}

\textcolor{red}{In order to identify the SNR regime for which the multiscale methods are reliable, the F1-scores of multiscale GSPWD and multiscale SPWD were averaged for each scene with $N>1$, and an rmANOVA was run with \textit{SNR} as the within-subjects factor and \textit{SH Order} and \textit{Number of Sources} as the between-subjects factors, respectively. Since the sphericity assumption was violated (Mauchly's test, $p < 0.001$), Greenhouse-Geisser correction was applied to the degrees of freedom of the within-subjects terms with $\varepsilon = 0.458$. \textit{SNR} was statistically significant with a large effect size [$F(1.831, 4486.062) = 7631.500$, $p < 0.001$, $\omega^2 = 0.463$], while its interactions with \textit{SH Order} [$F(7.324, 4486.062) = 288.572$, $p < 0.001$, $\omega^2 = 0.066$] and \textit{Number of Sources} [$F(16.479, 4486.062) = 185.469$, $p < 0.001$, $\omega^2 = 0.055$] had medium and small effect sizes, respectively. The three-way interaction had a negligible effect size. Post-hoc comparisons using Bonferroni correction indicated that the mean F1-score at $10$ dB SNR was significantly lower than all other levels (mean differences between $0.239$ and $0.243$, $p < 0.001$). Among the remaining levels, only the difference between $20$ and $30$ dB SNR was statistically significant ($p = 0.016$). However, all pairwise differences among $20$, $30$, $40$ dB SNR and the noise-free condition were at most $0.004$ in F1-score and therefore practically negligible. This indicates that the performance of the multiscale methods is stable when SNR is $20$ dB or higher.}


We marginalized F1-scores over $20$, $30$, $40$ dB \textit{SNR} and the noise-free conditions to evaluate the methods in the operating regime of the proposed multiscale methods. Fig.~\ref{fig:f1marginal} shows the F1-scores {averaged over \textit{Number of Sources} and \textit{SNR}} with respect to  \textit{SH Order}. It may be observed that average F1-scores for the multiscale GSPWD and SPWD methods increase monotonically with SH order and both exceed $0.9$ at $N=6$ indicating that precision and recall are simultaneously high. In contrast, average F1-scores for single GSPWD and SPWD decrease with increasing SH order beyond $N=4$, showing substantial degradation at higher SH orders. 

In order to compare the methods at higher SH orders in the same operating range, the F1-scores of each scene were averaged over the SNR levels of $20$ dB and above, and separate rmANOVAs with \textit{Method} as the within-subjects factor and \textit{Number of Sources} as the between-subjects factor were run for $N = 4$, $5$ and $6$. Table~\ref{tab:contrasts_f1} shows the post-hoc comparisons of multiscale GSPWD with each of the other methods, where positive estimates indicate a higher F1-score for multiscale GSPWD.

\begin{table}[t]
\centering
\caption{Post-hoc comparisons of F1-score (Multiscale GSPWD $-$ Other Method) at $N = 4$, $5$, and $6$ for $\textrm{SNR} \geq 20$~dB. $p$-values are Bonferroni-corrected within each SH order.}
\label{tab:contrasts_f1}
\begin{tabular}{llcccc}
\toprule
$N$ & Comparison & Estimate & $t(490)$ & $p$ & Cohen's $d$ \\
\midrule
\multirow{3}{*}{4}
  & SPWD (Multiscale) & 0.003 & 0.625  & 1.000    & 0.030 \\
  & GSPWD (Single)    & 0.018 & 3.399  & 0.004    & 0.180 \\
  & SPWD (Single)     & 0.006 & 0.990  & 1.000    & 0.058 \\
\midrule
\multirow{3}{*}{5}
  & SPWD (Multiscale) & 0.020 & 4.329  & $< .001$ & 0.194 \\
  & GSPWD (Single)    & 0.064 & 10.895 & $< .001$ & 0.616 \\
  & SPWD (Single)     & 0.090 & 14.722 & $< .001$ & 0.861 \\
\midrule
\multirow{3}{*}{6}
  & SPWD (Multiscale) & 0.010 & 2.328  & 0.122    & 0.090 \\
  & GSPWD (Single)    & 0.161 & 22.158 & $< .001$ & 1.501 \\
  & SPWD (Single)     & 0.246 & 36.871 & $< .001$ & 2.297 \\
\bottomrule
\end{tabular}
\end{table}

\textcolor{red}{It may be observed that the advantage of using multiscale GSPWD in terms of F1-score emerges particularly for $N>4$ where multiscale GSPWD also provides a statistically significant improvement over single scale methods where the effect size is largest ($d = 2.297$) between multiscale GSPWD and single SPWD at $N=6$. These results also indicate that multiscale analysis is increasingly essential for higher SH orders, with multiscale GSPWD providing the best average F1-score performance.}

\subsubsection{Event Count Mismatch (ECM)}

\begin{figure*}[!t]
    \centering
    \includegraphics[width=0.85\linewidth]{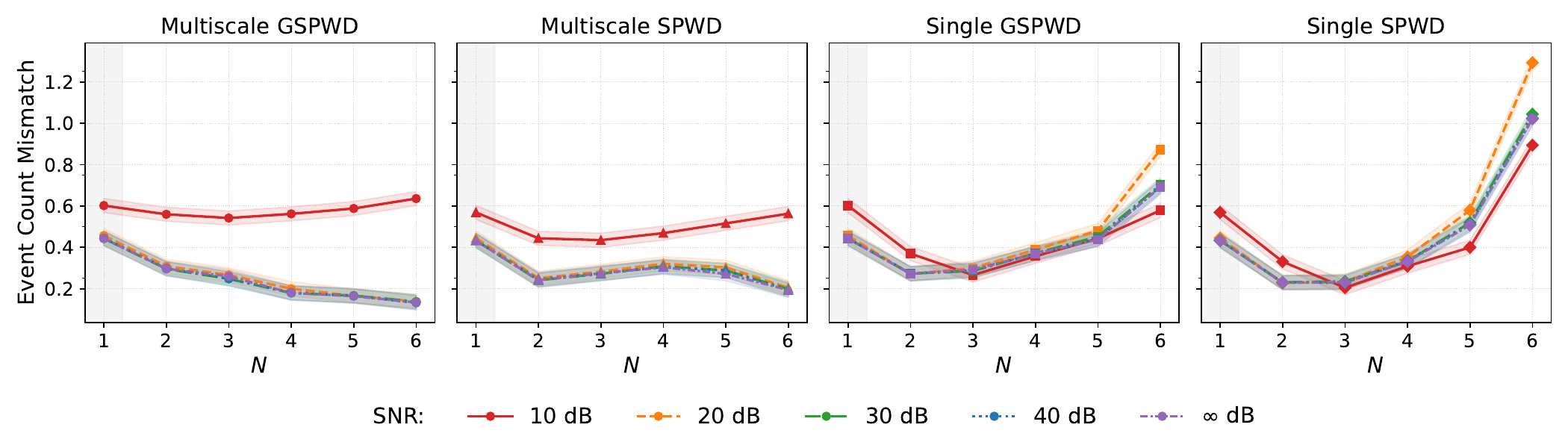}
    \caption{Event Count Mismatch (ECM) for different methods and SNR levels. The curves represent the mean values and shaded areas represent the 95\% CI above and below the mean. $N=1$ for which multiscale versions collapse into their single scale versions is indicated by gray shading.}
    \label{fig:ecm_marginals}
\end{figure*}

\begin{figure*}[!t]
    \centering
    \includegraphics[width=0.85\linewidth]{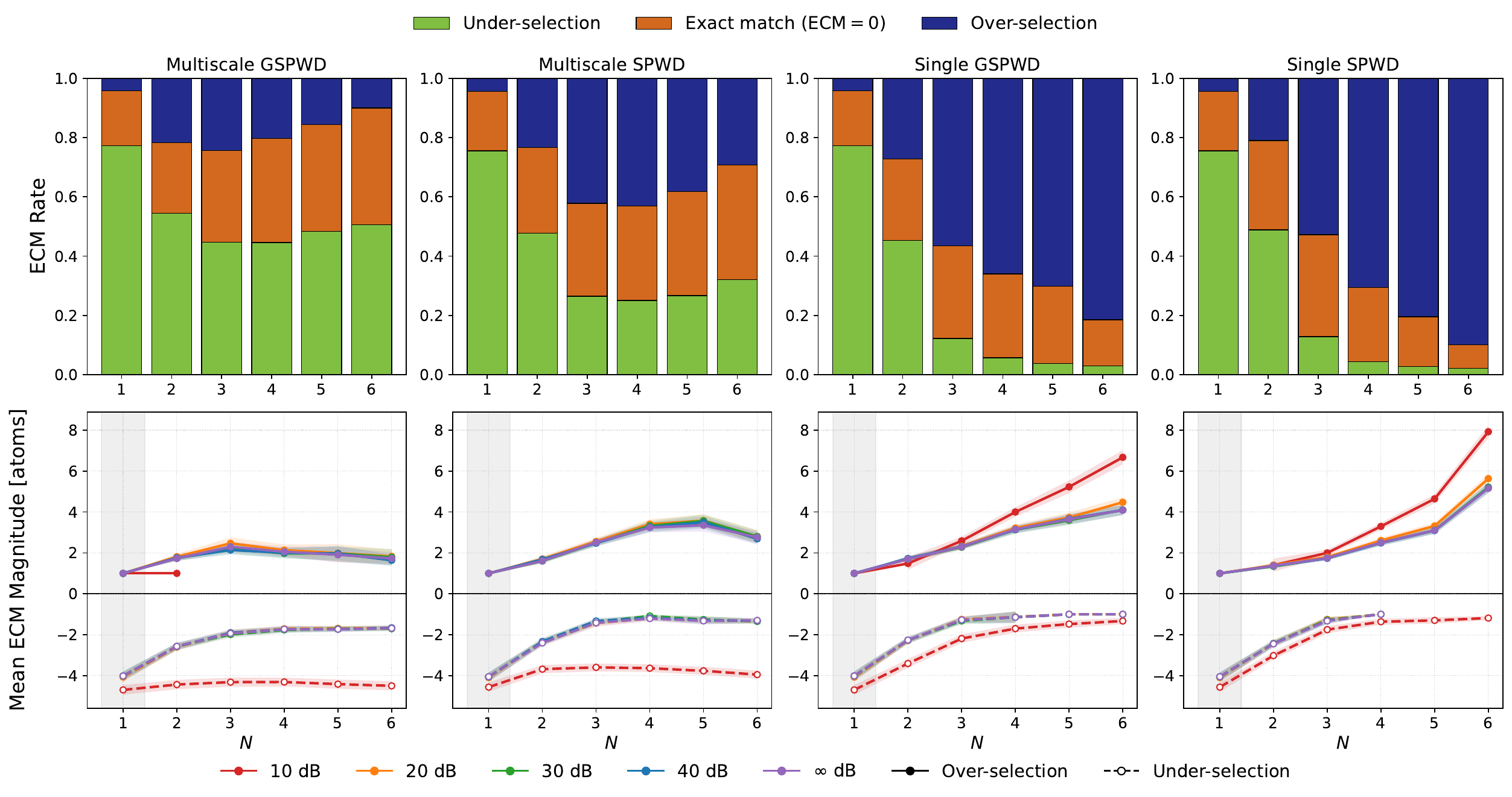}
    \caption{Event Count Mismatch (ECM) Rate (top row) and ECM Magnitude (bottom row) for different methods and SNR levels. 95\% CIs for Mean ECM Magnitudes are also shown for each noise level. $N=1$ for which multiscale versions collapse into their single scale versions is indicated by gray shading.}
    \label{fig:ecm_rate_magnitude}
\end{figure*}

\textcolor{red}{An rmANOVA of ECM with \textit{Method} and \textit{SNR} as within-subjects and \textit{SH Order} and \textit{Number of Sources} as between-subjects factors was carried out. Since the sphericity assumption was not satisfied for \textit{Method}, \textit{SNR} and their interaction (Mauchly's test, $p < 0.001$), Greenhouse-Geisser correction was applied to the degrees of freedom of all within-subjects terms with $\varepsilon = 0.775$, $0.399$ and $0.291$, respectively.}

\textcolor{red}{All main effects and interactions were statistically significant ($p < 0.001$). \textit{SNR} [$F(1.597, 4696.365) = 4992.296$, $\omega^2 = 0.280$], \textit{Method} [$F(2.324, 6833.242) = 2123.098$, $\omega^2 = 0.248$] and \textit{SH Order} [$F(5, 2940) = 802.108$, $\omega^2 = 0.185$] had large effect sizes, while \textit{Number of Sources} [$F(9, 2940) = 79.902$, $\omega^2 = 0.024$] had a small effect size. Among the interactions, \textit{Method} $\times$ \textit{SH Order} [$F(11.621, 6833.242) = 431.576$, $\omega^2 = 0.106$] had a medium effect size, indicating that the methods have different ECM performances for different SH orders. \textit{Method} $\times$ \textit{SNR} [$F(3.488, 10253.925) = 468.332$, $\omega^2 = 0.053$], \textit{SNR} $\times$ \textit{Number of Sources} [$F(14.377, 4696.365) = 73.630$, $\omega^2 = 0.018$], \textit{Method} $\times$ \textit{SNR} $\times$ \textit{Number of Sources} [$F(31.390, 10253.925) = 55.644$, $\omega^2 = 0.015$], \textit{SNR} $\times$ \textit{SH Order} [$F(7.987, 4696.365) = 61.192$, $\omega^2 = 0.012$] and \textit{Method} $\times$ \textit{SNR} $\times$ \textit{SH Order} [$F(17.439, 10253.925) = 39.954$, $\omega^2 = 0.010$] had small effect sizes, while the remaining interactions were negligible.}

Fig.~\ref{fig:ecm_marginals} shows the ECM marginalized over \textit{Number of Sources} for different levels of \textit{SNR}. It may be observed that multiscale methods perform well at $\textrm{SNR}\geq{}20$ dB, where both achieve the low ECMs of $0.14$ and $0.194$ at $N=6$ for multiscale GSPWD and multiscale SPWD, respectively. The single scale methods have monotonically decreasing ECM performance for $N\geq{}3$. The susceptibility of multiscale methods to noise can also be observed.

{Fig.~\ref{fig:ecm_rate_magnitude} shows the ECM Rate and ECM Magnitude for different methods. ECM Rate (top row) indicates that multiscale methods provide a higher proportion of exact matches (i.e. $|\mathcal{S}|=|\mathcal{Q}|$) than the single scale methods with comparable performance among them, whereas single scale methods more frequently over-select, particularly at higher SH orders. Multiscale GSPWD under-selects more often than it over-selects. However, both over- and under-selection are below approximately 2 atoms for multiscale GSPWD where a higher over-selection is observed for multiscale SPWD.}

\begin{figure*}[!th]
    \centering
    \includegraphics[width=.85\linewidth]{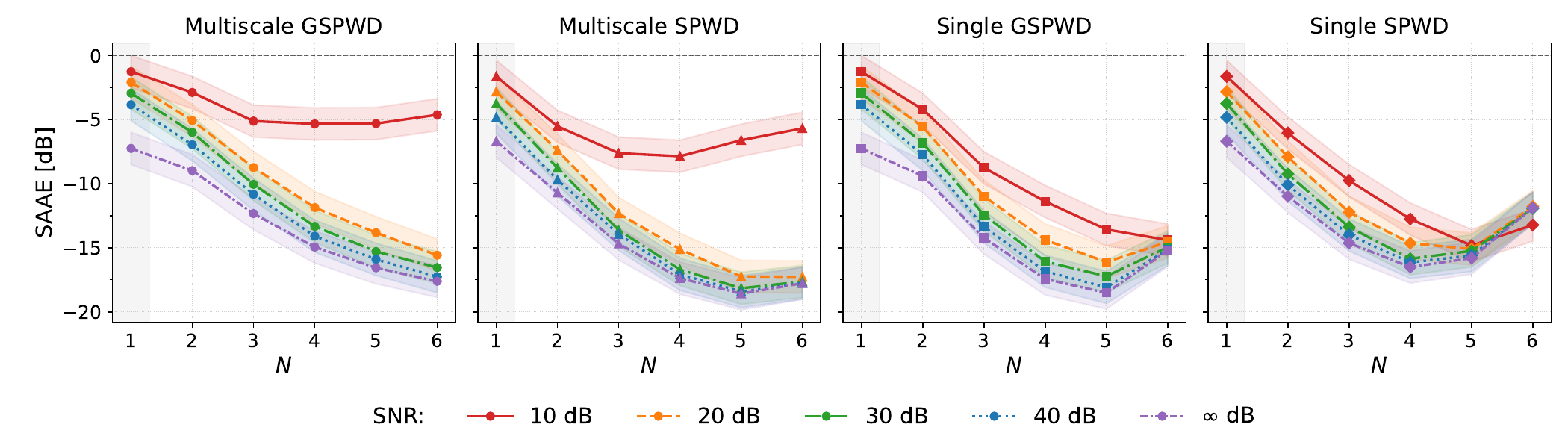}
    \caption{Support-aware Amplitude Error (SAAE) for different methods, SNR levels and SH orders. $N=1$ for which multiscale versions collapse into their single scale versions is indicated by gray shading.}
    \label{fig:saae}
\end{figure*}

{\subsubsection{Support-aware Amplitude Error (SAAE)}

\textcolor{red}{An rmANOVA of SAAE with \textit{Method} and \textit{SNR} as within-subjects and \textit{SH Order} and \textit{Number of Sources} as between-subjects factors was carried out. Since the sphericity assumption was not satisfied for \textit{Method}, \textit{SNR} and their interaction (Mauchly's test, $p < 0.001$), Greenhouse-Geisser correction was applied to the degrees of freedom of all within-subjects terms with $\varepsilon = 0.906$, $0.563$ and $0.535$, respectively.}

\textcolor{red}{All main effects and interactions were statistically significant ($p < 0.001$). \textit{Number of Sources} [$F(9, 2940) = 1992.579$, $\omega^2 = 0.379$] and \textit{SNR} [$F(2.253, 6622.986) = 2302.321$, $\omega^2 = 0.144$] had large effect sizes, \textit{SH Order} [$F(5, 2940) = 472.042$, $\omega^2 = 0.118$] had a medium effect size and \textit{Method} [$F(2.717, 7986.744) = 190.232$, $\omega^2 = 0.019$] had a small effect size. Among the interactions, \textit{SNR} $\times$ \textit{Number of Sources} [$F(20.274, 6622.986) = 505.668$, $\omega^2 = 0.109$] had a medium effect size, while \textit{Method} $\times$ \textit{SNR} [$F(6.415, 18858.767) = 440.493$, $\omega^2 = 0.032$], \textit{Method} $\times$ \textit{Number of Sources} [$F(24.449, 7986.744) = 70.598$, $\omega^2 = 0.019$], \textit{Method} $\times$ \textit{SNR} $\times$ \textit{SH Order} [$F(32.073, 18858.767) = 84.328$, $\omega^2 = 0.016$], \textit{SNR} $\times$ \textit{SH Order} $\times$ \textit{Number of Sources} [$F(101.372, 6622.986) = 42.100$, $\omega^2 = 0.013$], \textit{Method} $\times$ \textit{SH Order} [$F(13.583, 7986.744) = 42.846$, $\omega^2 = 0.010$] and \textit{SH Order} $\times$ \textit{Number of Sources} [$F(45, 2940) = 31.260$, $\omega^2 = 0.010$] had small effect sizes. The remaining interactions fell below the $\omega^2 \geq 0.01$ threshold for practical significance.}

{Post-hoc comparisons with Bonferroni correction indicate that multiscale GSPWD has significantly higher SAAE than the other methods at $p<0.001$ level with mean differences ranging from $1.63$ to $2.03$ dB and small-to-medium effect sizes ($d = 0.24–0.30$) indicating comparatively lower amplitude estimation accuracy.}

{Fig.~\ref{fig:saae} shows SAAE averaged over source count for different methods at different SNR levels and different SH orders. It may be observed that the SAAE performance of all methods degrade with increasing noise. Multiscale methods perform worse than single scale methods at 10 dB SNR, but their performances improve with increasing SNR for the other tested noise levels. An interesting observation is that performance of single-scale methods start to degrade at higher SH orders (i.e.~$N>4$) in a way similar to other measures.}

\subsubsection{Localization Error (LE)}
\begin{figure}
    \centering
    \includegraphics[width=0.4\textwidth]{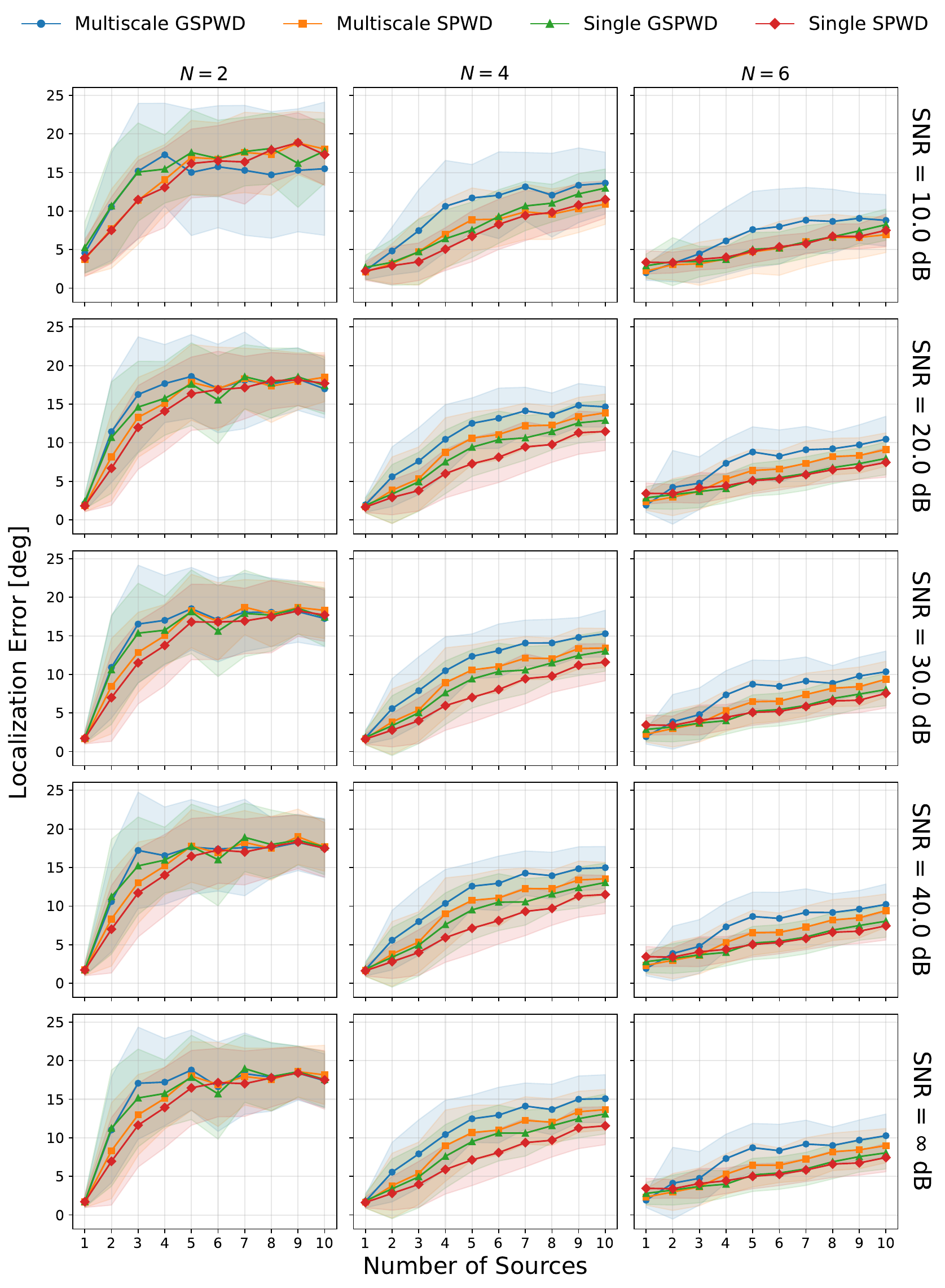}
    \caption{Localization Error (LE) in degrees for different methods and number of sources. Only the conditions for $N=2$, $N=4$ and $N=6$ are shown for conciseness. The curves represent the mean values and shaded areas represent one standard deviation above and below the mean.}
    \label{fig:le}
\end{figure}

Part of the localization error (LE) results for the tested methods are shown in Fig.~\ref{fig:le}. It may be observed that the localization error performance is similar across different methods. LE increases in general with the number of sources with almost perfect localization (i.e. less than the dictionary resolution) for a single plane wave irrespective of SNR. Similarly, increasing SH order improves the LE for all the tested methods.

\textcolor{red}{An rmANOVA of LE with \textit{Method} and \textit{SNR} as within-subjects and \textit{SH Order} and \textit{Number of Sources} as between-subjects factors was carried out. Since the sphericity assumption was not satisfied for \textit{Method}, \textit{SNR} and their interaction (Mauchly's test, $p < 0.001$), Greenhouse-Geisser correction was applied to the degrees of freedom of all within-subjects terms with $\varepsilon = 0.645$, $0.464$ and $0.330$, respectively.}

\textcolor{red}{\textit{SH Order} [$F(5, 2940) = 495.407$, $p < 0.001$, $\omega^2 = 0.123$] and \textit{Number of Sources} [$F(9, 2940) = 323.620$, $p < 0.001$, $\omega^2 = 0.090$] had medium effect sizes, and \textit{Method} [$F(1.935, 5689.174) = 273.334$, $p < 0.001$, $\omega^2 = 0.032$] had a small effect size. \textit{SNR} [$F(1.854, 5450.869) = 86.075$, $p < 0.001$, $\omega^2 = 0.005$] was statistically significant but with negligible effect size. All significant interactions had effect sizes below $\omega^2 = 0.01$ indicating a negligible magnitude. In particular, the negligible \textit{Method} $\times$ \textit{SH Order} interaction ($\omega^2 = 0.007$) indicates that the differences between the methods in LE do not depend substantially on the SH order.}

\textcolor{red}{A post-hoc comparison for \textit{Method} $\times$ \textit{SH Order} with Bonferroni correction indicated that for $N \geq 3$, multiscale GSPWD had a significantly higher LE than the other methods ($p < 0.001$), with mean differences between $1.43^{\circ}$ and $3.67^{\circ}$ and small-to-medium effect sizes ($0.29 \leq d \leq 0.74$). The largest differences were observed with respect to single SPWD at $N=4$ ($MD = 3.67^{\circ}$, $t(2940) = 15.555$, $p < 0.001$, $d = 0.742$) and $N=5$ ($MD = 3.61^{\circ}$, $t(2940) = 15.333$, $p < 0.001$, $d = 0.731$). At $N=2$, multiscale GSPWD differed significantly only from single SPWD ($MD = 1.15^{\circ}$, $p < 0.001$, $d = 0.233$). These differences are significantly less than the angular resolution of the GSRF map (i.e.~$29.32^{\circ}$). Therefore, it can be stated that despite the multiscale versions providing an objectively worse LE performance, all of the compared methods provide a similar localization accuracy in practical terms.}


{\subsection{Effects of Dictionary Resolution}}
{Another set of simulations was run with $10$ repetitions per scenario to analyze the effect of dictionary resolution on performance. The compared dictionary sizes were $L=48$, $L=192$, $L=768$ and $L=3072$. Only the $\textrm{SNR}=\infty$ condition was tested for $N=1\cdots{}6$ and with $H=48$. The simulations were run on an Apple MacBook Pro M4 with 24 GB RAM using Python 3.14. Fig.~\ref{fig:L-all-metrics} indicate the performance using the measures described earlier, as well as the average compute time, marginalized over the \textit{SH Order} and the \textit{Number of Sources}.}

\begin{figure*}[t]
	\centering
	\subfigure[\label{fig:hf1}]{
		\includegraphics[width=0.188\textwidth]{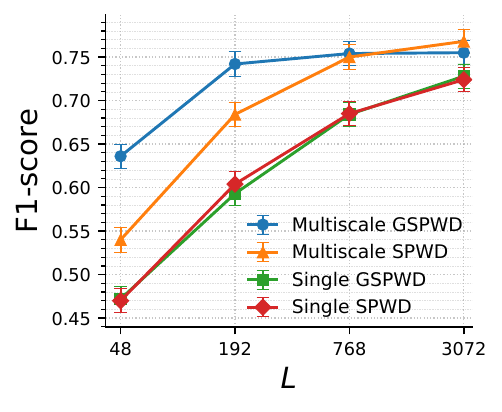}}
    \subfigure[\label{fig:hecm}]{
		\includegraphics[width=0.188\textwidth]{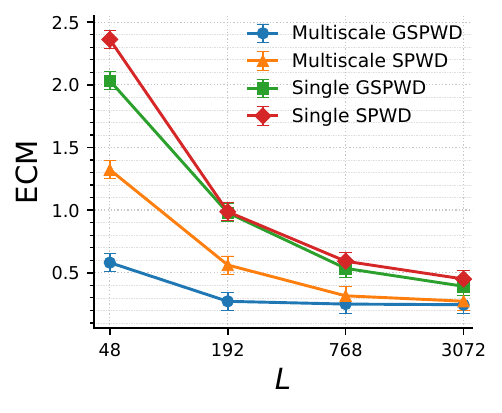}}
    \subfigure[\label{fig:hsaae}]{
        \includegraphics[width=0.188\textwidth]{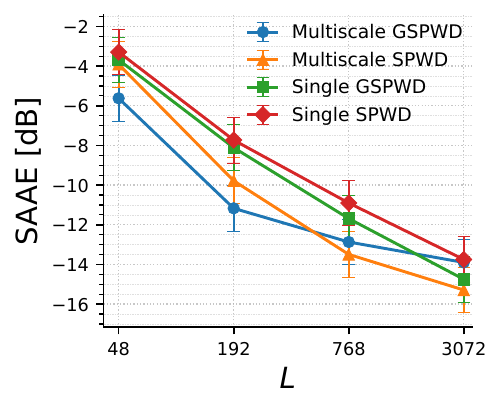}}
    \subfigure[\label{fig:hle}]{
		\includegraphics[width=0.188\textwidth]{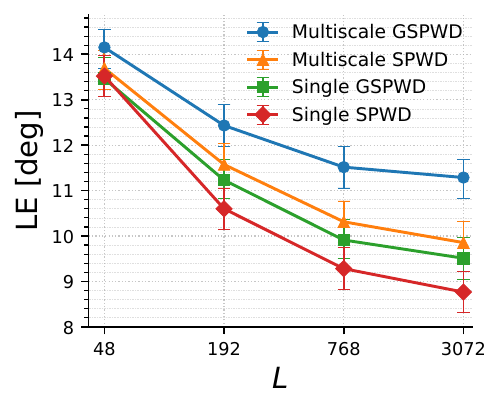}}
	\subfigure[\label{fig:ltime}]{
        \includegraphics[width=0.188\textwidth]{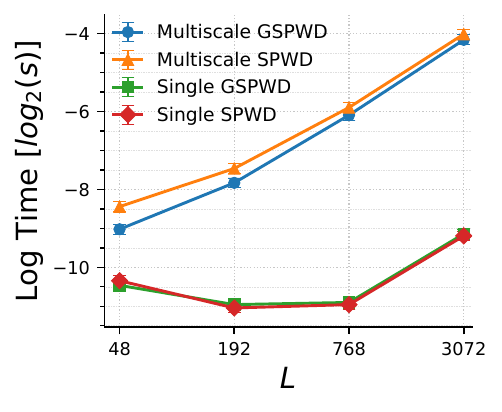}}	%
	\caption{Marginal means (with 95\% CI) of F1-score, ECM, SAAE, LE, and log-runtime with respect to $L$.}
	\label{fig:L-all-metrics}
\end{figure*}

{Fig.~\ref{fig:hf1} shows that increasing the dictionary resolution increases the F1-scores for all methods, while multiscale methods provide a higher mean F1-score even when the dictionary resolution is much lower at $14.7^{\circ}$ for $L=192$. Overall, multiscale methods perform consistently better than the single scale versions regardless of the dictionary resolution. }

{Fig.~\ref{fig:hecm} indicates that multiscale GSPWD provides the best performance even at the low resolution achieving an ECM of $0.2$ at $L=192$. The same ECM is achieved by multiscale SPWD only at the much higher resolution afforded by selecting $L=3072$.}

{Fig.~\ref{fig:hsaae} shows that SAAE decreases with increasing $L$, for all methods, which is to be expected since the mismatch between the dictionary and continuously sampled directions of the plane waves in the simulation gets smaller with increasing resolution. As with the F1-score and ECM, multiscale GSPWD provides a sufficient performance for $L=192$, with less of an improvement than the other methods beyond that resolution.}

{Fig.~\ref{fig:hle} indicates that LE for all methods decrease with increasing $L$. However, a low LE can be obtained for single scale versions subject to the caveat associated with LE discussed earlier. In all cases, LE is less than half the angular resolution of SRF/GSRF.}

{While a formal analysis of computational complexity is beyond the scope of this article, Fig.~\ref{fig:ltime} shows the effect of the directional resolution of the dictionary on compute time. The major factor that affects the computational requirements is the resolution of the dictionary which determines $L$ and the algorithm type. It may be observed that increasing dictionary resolution exponentially results in a similar exponential increase in mean compute time per SRF/GSRF.}

\subsection{Discussion}

Both multiscale approaches provide F1-scores that are comparable with single scale methods up to $N=4$. For $N\geq{}4$, the mean F1-scores for single scale algorithms decrease while those of multiscale algorithms continue to monotonically increase, with multiscale methods reaching above $0.9$ at $N=6$ indicating improved performance simultaneously for precision and recall. \textcolor{red}{The statistical analysis indicated that this performance is largely independent of noise when $\textrm{SNR}\geq{}20$.}

The proposed multiscale methods have an average support recovery performances that are better that those of single-scale methods, particularly at medium and low noise levels (i.e. $\textrm{SNR}\leq{}20$) and higher SH orders. \textcolor{red}{While single SPWD and single GSPWD have increasing event count mismatch at high SH orders (indicating that these methods will identify spurious plane waves that are not in the ground truth), both multiscale versions, and in particular multiscale GSPWD, provide a higher performance when $\textrm{SNR}\geq{}20$.}

{While multiscale GSPWD showed the highest support recovery accuracy, this came at the cost of higher amplitude estimation error relative to the other three methods. The amplitude recovery performance measured by SAAE shows that the proposed multiscale GSPWD performs slightly worse than the other methods.}

Multiscale GSPWD method provides a localization accuracy that is comparable to if slightly worse than those of single scale methods. However, the average localization error is less than half the grid resolution of the SRF map indicating limited adverse effects on performance. {We speculate that the improved LE performance of single scale methods is caused by overselection that these methods provide, overfitting the underlying measurement vector with spurious components.}

{Although the OMP iterations within a single recovery are sequential, block-based time-frequency processing produces one independent GSRF per time-frequency bin, which can be batched and solved concurrently on a GPU. This suggests a practical path to real-time operation even though the present implementation processes each instance sequentially on the CPU and is non-optimal due to its reliance on an interpreted language which is inherently slow.}

\section{Conclusions}\label{sec:conc}

Two improvements to SPWD were presented in this article. The first improvement targets optimizing the employed spherical kernel by trading off the mainlobe half-width for reducing the sidelobes of the spherical kernel and eliminating the backlobe, thereby reducing the off-band correlation of kernels, particularly in the direction of the backlobe. SPWD when carried out with this kernel is called generalized SPWD (GSPWD). The second improvement is a multiscale extension of the GSPWD which is based on the observation that SRFs at different SH orders nominally have the same sparsity profile and the same activations when projected on dictionaries of spherical kernels calculated at different orders. This allows the development of a sparse recovery method that can provide improved performance in terms of sparse recovery performance.

A numerical simulation to assess the performance of the proposed method indicated that while the proposed multiscale approach has slightly worse localization error, it has an improved sparse recovery performance in comparison with the single SPWD (baseline), single GSPWD and multiscale SPWD.

While the empirical evidence demonstrated via simulations indicate that the proposed approach performs well in practice in terms of accuracy and support recovery, a study of theoretical exact recovery guarantees is left for future work. Also, stopping criterion depending on SNR which can potentially improve support recovery performance further is left for a future investigation.  Similarly, the proposed approach is empirically shown to be robust to isotropic diffuse noise for $\textrm{SNR}\leq{}20$ dB, however analyses of its performance under different noise profiles is also left for future work.




%

\bibliographystyle{IEEEtran} 
\bibliography{biblio} 


\vfill

\begin{IEEEbiography}[{\includegraphics[width=1in,height=1.25in,clip,keepaspectratio]{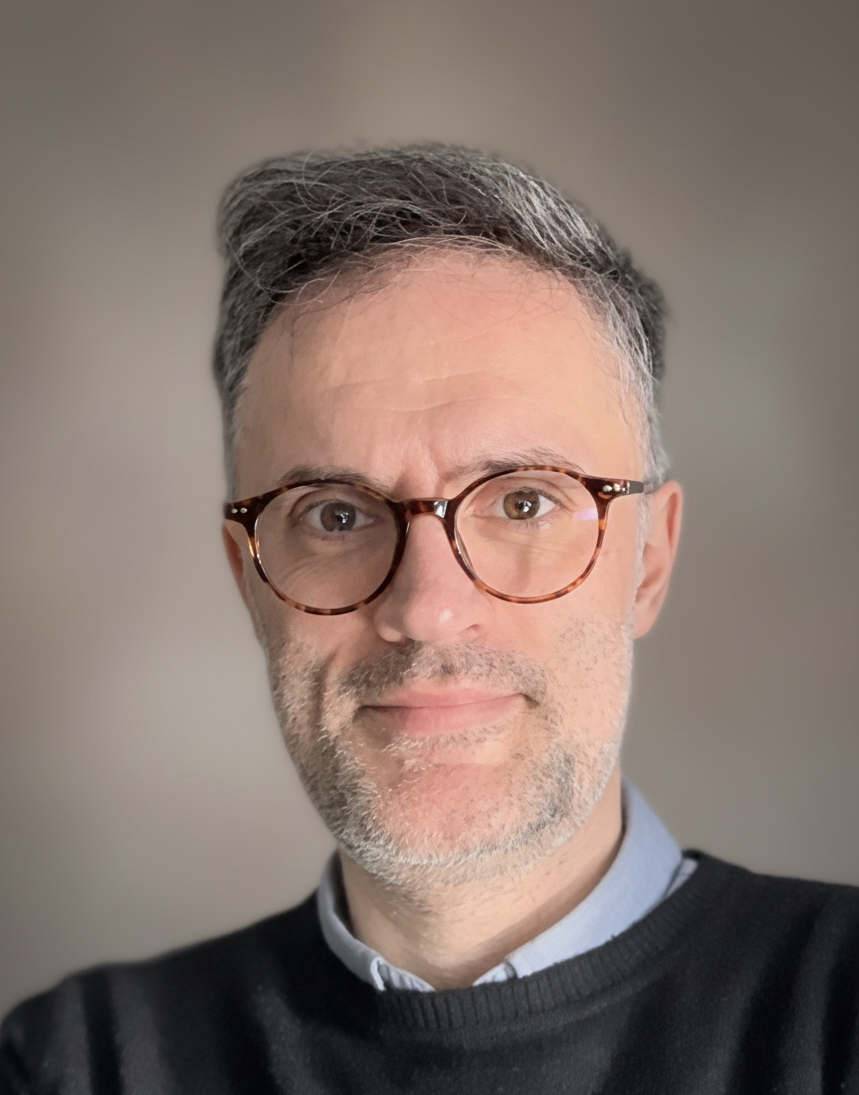}}]{H\"{u}seyin Hac\i{}habibo\u{g}lu} (Senior Member, IEEE) received the B.Sc. (Hons.) degree in electrical and electronic engineering from Middle East Technical University (METU), Ankara, Türkiye, in 2000, the M.Sc. degree in electrical and electronic engineering from the University of Bristol, Bristol, U.K., in 2001, and the Ph.D. degree in computer science from Queen’s University Belfast, Belfast, U.K., in 2004. He held research positions with the University of Surrey, Guildford, U.K., from 2004 to 2008, and King’s College London, London, U.K., from 2008 to 2011. He is currently a Professor of signal processing with the Graduate School of Informatics, METU. His research interests include audio signal processing, room acoustics, immersive audio systems, psychoacoustics of spatial hearing, and microphone arrays. Prof. Hac\i{}habibo\u{g}lu is a Member of the IEEE Signal Processing Society, Audio Engineering Society (AES), Turkish Acoustics Society (TAD), European Acoustics Association (EAA), and Acoustical Society of America (ASA). He was an Associate Editor for IEEE/ACM TRANSACTIONS ON AUDIO, SPEECH AND LANGUAGE PROCESSING between 2017-2021 and IEEE SIGNAL PROCESSING LETTERS between 2021-2025. He is currently an official representative of METU in Moving Picture, Audio and Data Coding by Artificial Intelligence (MPAI) and Audio Engineering Society (AES) Vice President for the Southern Europe, Middle East and Africa (SEMEA) region.
\end{IEEEbiography}

\end{document}